# Reproducibility and Leakage-Controlled Evaluation of Machine-Learning Models for Chronic Kidney Disease Classification

**Lening Yang, Yichuan Zhao**

## Abstract

**Background:** Machine-learning studies based on small clinical datasets can report near-perfect classification, yet their apparent performance may be highly sensitive to preprocessing, data leakage, missing-data handling, and the choice of evaluation sample. We reproduced a published chronic kidney disease (CKD) classification study and broadened its evaluation by introducing explicit information boundaries, repeated cross-validation, calibration assessment, and model interpretation.

**Methods:** The public UCI CKD dataset contains 400 records, 24 predictors, and a binary diagnostic label. Before analysis, we audited strings, variable types, class labels, missing-value markers, and column alignment. CKD was encoded as 1 and non-CKD as 0. A stratified 70:30 development–holdout partition was created with random seed 12 before any learned preprocessing. Because the labels in this public holdout had been seen during earlier development, we treated it as a locked historical internal sample rather than as confirmatory validation. Imputation, one-hot encoding, scaling, supervised feature selection, and principal component analysis (PCA) were fitted only within development-set training folds. We used two cross-validation designs: a single five-fold stratified cross-validation to compare our findings with the published XGBoost results, and five-fold stratified cross-validation repeated three times for the main comparison of nine classifiers across three feature representations and three missing-data strategies. Models were ranked solely on training cross-validation, and exact ties were reported as jointly top-ranked. A frozen set of finalists was then evaluated once on the holdout for discrimination, calibration, Brier score, and conditional stratified-bootstrap percentile intervals. Model behavior was further examined through leave-one-feature-out ablation, an ordinal-encoding sensitivity analysis, and SHAP values grouped by original variable. To assess dependence on the random partition, we also repeated the entire partitioning, cross-validation, selection, and holdout procedure from scratch under three additional random seeds (42, 39, and 55). After the main analysis had been frozen, we added post hoc sensitivity analyses on the persisted development folds (three of them also drew on records outside the main development sample): nested resampling of the complete selection procedure, characterization of the missingness pattern, multiple imputation, feature-selection stability, removal of strongly diagnostic predictors, uniform class weighting across classifiers, and a comparison with a naive analysis that fitted preprocessing on all records.

**Results:** The audit retained 250 CKD and 150 non-CKD records and repaired one shifted row. Among approaches that retained all records, selected original features with numeric *k*-nearest-neighbor imputation and Extra Trees served as the display representative of the top-ranked selected-feature configuration. It achieved a repeated-cross-validation accuracy of 0.9869 (standard deviation, 0.0218), ROC area of 1.0000, and holdout accuracy of 0.9833 (exact 95% interval, 0.9411–0.9980). Of the representative retained-sample models, the PCA Random Forest pipeline had the lowest holdout Brier score (0.0053). PCA retained 27 components that explained 95.96% of the variance and performed similarly to the best original-feature representation. Complete-case deletion reduced the development and holdout samples to 112 and 46 records, respectively, and reversed the majority class. Eight configurations were tied at the top of that branch, whose display representative achieved holdout accuracy of 1.0000 (46 of 46; exact 95% interval, 0.9229–1.0000). Removing specific gravity produced the largest mean paired ablation loss (−0.0238), although the effect of removing any single field could not be distinguished from fold-to-fold variation. Hypertension made the largest grouped SHAP contribution. Across the three additional seeds, the selected-feature pipeline from the main analysis remained jointly top-ranked in two replications. Nested resampling of the entire selection procedure yielded an outer accuracy of 0.9845 (standard deviation, 0.0177), 0.0095 below the same-fold accuracy of the top-ranked retained-sample pipeline (PCA Random Forest), and the selected pipeline varied across outer folds. Missingness was itself informative: classifiers given only the 24 missing-value indicators reached a ROC area of 0.85–0.88. Multiple imputation, uniform class weighting, and naive global preprocessing each changed mean accuracy by less than 0.01, whereas removing the renal, urinary, and hematologic fields reduced accuracy to 0.89, and seven demographic, metabolic, and comorbidity fields alone yielded 0.84.

**Conclusions:** High internal performance persisted after leakage controls were imposed. PCA raised the cross-validation accuracy of XGBoost, the classifier compared in the source study, by 0.0155. Compared with the best interpretable original-feature representations, however, PCA produced only a marginal cross-validation difference in accuracy (0.0036, computed before rounding); this difference fell within fold-to-fold variation and did not yield a consistent holdout advantage across random seeds. Several distinct pipelines were also statistically indistinguishable. The near-perfect discrimination appears to be driven largely by characteristics of the dataset, particularly clinically proximal laboratory and urinalysis variables, with CKD-associated missingness as an additional informative signal. Near-perfect estimates from a small, single-source dataset do not amount to clinical validation. External and prospective evaluation remains necessary before clinical use.

## 1. Introduction

Chronic kidney disease (CKD) is a major contributor to global morbidity and mortality, and its health burden has grown over recent decades [1]. Current clinical guidance emphasizes evaluation through measures of kidney function and kidney damage, followed by risk-based monitoring and management [2]. Computational prediction may help clinicians organize multidimensional information, but a model's apparent performance depends as much on study design and validation as it does on the learning algorithm.

Islam et al. evaluated multiple machine-learning algorithms on the UCI CKD dataset and reported high test accuracy, including approximately 0.983 for XGBoost on the original data and 0.9916 for XGBoost in an analysis described as PCA based [3]. These results are noteworthy because the dataset is openly available [4] and widely used for teaching and benchmarking, yet exact reproduction is difficult. The report discusses several missing-data procedures, presents named original variables in its account of PCA, and omits some implementation details needed for exact replication. These ambiguities are consequential: fitting transformations to the full dataset before resampling can allow information from validation observations to enter model training and thereby inflate performance estimates [5].

Frameworks for prediction-model reporting and risk-of-bias assessment stress transparent data flow, appropriate missing-data handling, separation of model development from evaluation, and reporting of both discrimination and calibration [6,7]. Small datasets pose further problems. Performance can depend heavily on a favorable partition, observations recur across repeated cross-validation fits, exact ties between pipelines are common, and uncertainty may remain substantial even when average accuracy is high [8,9]. A rigorous reproduction should therefore document the source data, separate deterministic cleaning from learned preprocessing, quantify uncertainty, acknowledge ties instead of declaring nominal winners, and avoid treating internal validation as evidence of clinical transportability.

This study had three objectives. First, we reproduced the published comparison using the public UCI CKD dataset while auditing the source schema and ensuring that all learned preprocessing was estimated from training data alone. Second, we compared three feature representations—all original predictors, PCA of the expanded encoded matrix, and supervised selection of named clinical fields—under three approaches to missing data. Third, we extended the evaluation through repeated stratified cross-validation, tie-aware model selection, a single frozen evaluation on a locked historical holdout with conditional bootstrap intervals, calibration, Brier score, paired leave-one-feature-out ablation, an ordinal-encoding sensitivity analysis, and SHAP interpretation grouped by original variable. Rather than tuning the analysis to reproduce the published figure of 99.16%, we asked whether high performance would persist under a transparent, leakage-controlled design. The contribution is therefore an auditable, leakage-controlled re-analysis that traces apparent performance to its sources rather than proposing a new CKD classifier.

## 2. Materials and Methods

### 2.1 Study design

This secondary computational study reproduced and critically evaluated the methods of a published analysis based on an existing public dataset. The design was informed by the reporting principles of TRIPOD+AI and the risk-of-bias domains of PROBAST+AI [6,7]. It was not a prospective diagnostic trial, a clinical-impact study, or an external validation. Figure 1 outlines the analytical sequence and shows how model development, the selection and freezing of finalists, and holdout evaluation were kept separate.

### 2.2 Data source, predictors, and outcome

We obtained the Chronic Kidney Disease dataset from the UCI Machine Learning Repository [4]. The source file comprised 400 rows, 24 predictor fields, one outcome field, and one additional spreadsheet column. In accordance with the UCI metadata, 11 predictors were treated as numeric: age, blood pressure, random blood glucose, blood urea, serum creatinine, sodium, potassium, hemoglobin, packed cell volume, white blood cell count, and red blood cell count. The remaining 13 predictors were treated as categorical: specific gravity, albumin, sugar, red blood cells, pus cells, pus cell clumps, bacteria, hypertension, diabetes mellitus, coronary artery disease, appetite, pedal edema, and anemia. In the main analysis, specific gravity, albumin, and sugar were retained as nominal categorical variables because the repository metadata defines them as nominal levels. As these fields are clinically graded, we also conducted the prespecified ordinal-encoding sensitivity analysis described in Section 2.11.

The outcome was normalized by folding the case and removing spaces, tabs, underscores, and hyphens. Labels corresponding to "ckd" were mapped to CKD = 1, while "notckd" or "nonckd" were mapped to non-CKD = 0. We did not infer outcome values from predictor measurements.

### 2.3 Data audit and deterministic cleaning

Column names and string cells were normalized in Unicode, converted to lower case, and stripped of leading and trailing whitespace; repeated internal whitespace was collapsed. Empty values, question marks, repeated question marks, "NA," "N/A," "none," "null," and "missing" were converted to missing values. Numeric fields were coerced to numeric type, while categorical values were checked against prespecified allowed levels. Values outside those levels were treated as missing.

One record was shifted by one column from the diabetes field onward, leaving a valid "notckd" label in the additional column. We repaired this row using a deterministic rule that required an invalid label in the nominal target position and a valid label in the overflow position. The additional column was then removed. After cleaning, no invalid categorical or outcome values remained, leaving 400 rows and 25 analytical columns, including the outcome.

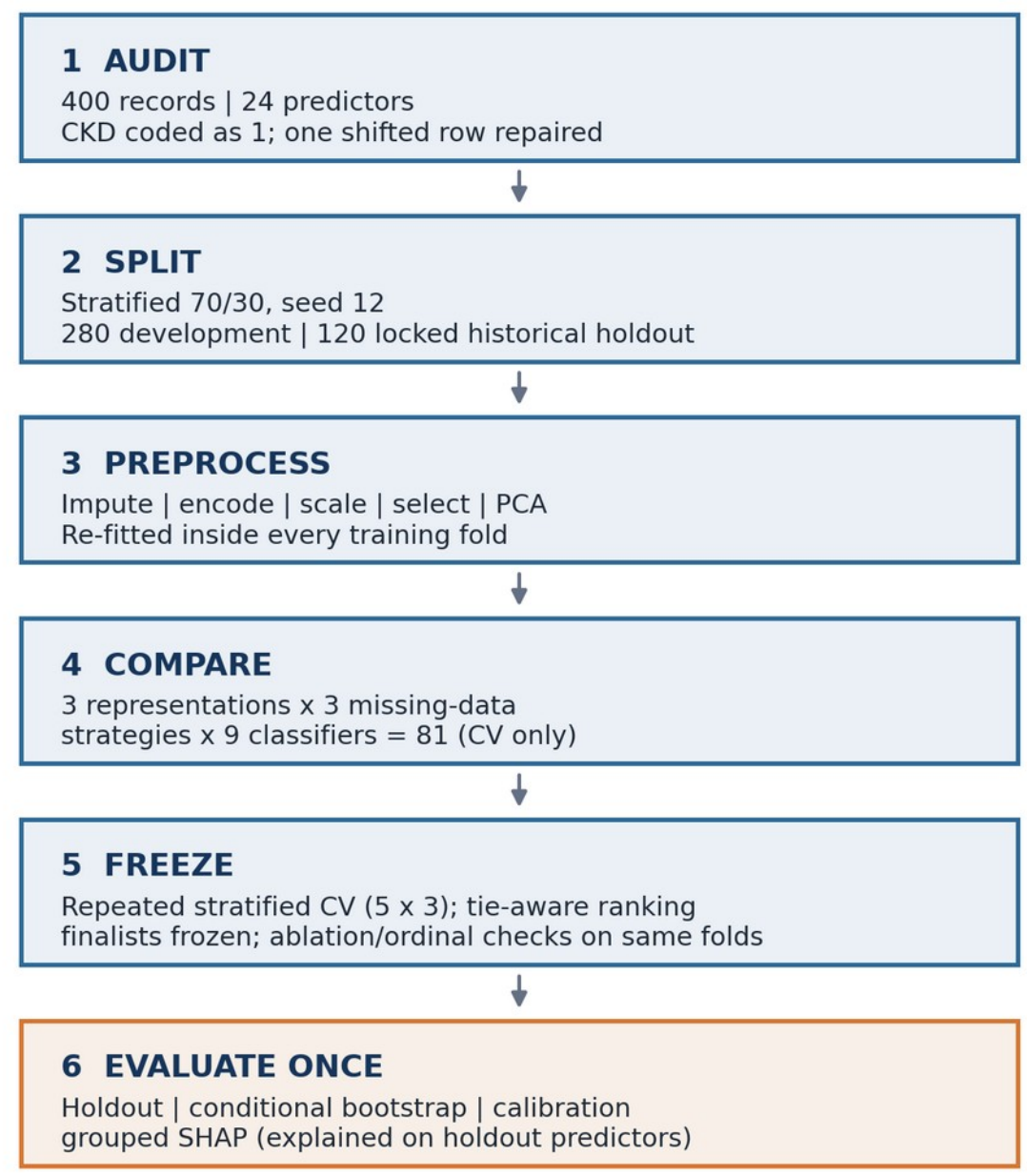

*Figure 1. Leakage-controlled analytical workflow. The holdout sample was separated before learned preprocessing or model selection. Imputation, encoding, scaling, feature selection, and PCA were re-estimated within each cross-validation training fold; finalists were frozen from training cross-validation before the holdout was evaluated once.*

### 2.4 Development and holdout samples

Before imputation, encoding, scaling, PCA, or supervised selection, we divided the data into a development sample (70%, $n$ = 280) and a holdout sample (30%, $n$ = 120) through stratified random sampling with seed 12. The development sample contained 175 CKD and 105 non-CKD records, while the holdout sample contained 75 CKD and 45 non-CKD records. We used the same partition throughout and retained the row assignments.

The labels in this public holdout had been visible during an earlier stage of the analysis. We therefore describe it as a locked historical internal holdout. Although it was excluded from every fitting, ranking, and selection step in the present analysis, it cannot be regarded as a new confirmatory or external sample. The earlier stage was an initial version of the same analysis, in which every record, including those now in the holdout, was used for model fitting or evaluation, although models were ranked by cross-validation only. The classifier settings (Section 2.7) and the rule fixing the selected panel at eight fields (Section 2.6) were already in place in that version and were not changed afterwards. None of these settings was chosen by comparing holdout results.

All learned transformations were incorporated into the modeling pipeline and re-estimated separately in every cross-validation training fold. Thus, validation-fold or holdout observations did not inform imputation values, category vocabularies, scaling parameters, principal axes, and supervised feature scores. Table 1 summarizes the experimental comparisons and their information boundaries.

### 2.5 Missing-data strategies

We compared three prespecified strategies. Under the median/mode strategy, numeric predictors were imputed with the training-fold median and categorical predictors with the most frequent training-fold category. Under the *k*-nearest-neighbor strategy, numeric predictors were first standardized using training-fold means and standard deviations, then imputed with a distance-weighted five-neighbor method [10]. This made neighbor distances comparable across variables whose raw scales differed by several orders of magnitude, such as white blood cell count and serum creatinine. Categorical predictors remained mode imputed to avoid imposing arbitrary distances between nominal levels. Under the complete-case strategy, rows with any missing predictor were deleted separately from the development and holdout samples after the initial split, with no subsequent imputation. Complete-case deletion does not estimate imputation parameters, but it can change outcome and covariate distributions when missingness is informative. We therefore treated it as a sensitivity analysis of a different analytical population rather than as the primary approach.

### 2.6 Feature representations

We evaluated three feature representations. The first included all 24 original clinical fields, with numeric scaling after imputation (or before *k*-nearest-neighbor imputation, as described above) and one-hot encoding of categorical variables. The second applied PCA after imputation, complete one-hot encoding, and standardization of the complete encoded matrix. Within each training fold, we retained the smallest number of components required to explain at least 95% of cumulative variance [11]. Since each expanded one-hot level receives unit variance, this procedure amounts to a level-weighted PCA of the expanded feature matrix rather than equal weighting of the 24 clinical variables. The third representation used supervised mutual information, estimated only within the relevant training fold, to retain eight of the 24 named clinical fields. The panel size, 30% of the predictors rounded up, followed the source study, which narrowed its variable list to 30% of the original variables [3]. For calculation of selection scores, missing values were temporarily filled inside the training fold with the same strategy as the downstream pipeline (median or scaled *k*-nearest-neighbor imputation for numeric fields and mode imputation for categorical fields; complete-case selection rejected missing values); the selector then passed the original selected columns to the pipeline, where preprocessing was re-estimated. For the *k*-nearest-neighbor strategy, selection scores used all candidate numeric variables to determine neighbor distances and thus provided a training-only screening approximation of the subsequent selected-panel imputation.

When fitted to the full development sample, the selected-feature display representative retained serum creatinine, sodium, hemoglobin, packed cell volume, red blood cell count, specific gravity, albumin, and hypertension. Because selection was repeated in every cross-validation fold, the retained fields could differ across validation iterations.

We treated PCA and original-feature selection as separate operations. PCA is an unsupervised rotation that produces linear combinations maximizing predictor variance, whereas mutual-information selection uses the training outcome and preserves clinically named measurements. We therefore evaluated their performance and interpretability separately.

### 2.7 Classifiers

Nine classifiers were prespecified: logistic regression, decision tree, Random Forest [12], support-vector machine (SVM), *k*-nearest neighbors, AdaBoost, gradient boosting, Extra Trees [13], and XGBoost [14]. Fixed configurations were used, with no hyperparameter selection based on a holdout sample. Logistic regression used the liblinear solver with a maximum of 5,000 iterations. The decision tree had a maximum depth of 6 and minimum leaf size of 2. Random Forest and Extra Trees each used 300 estimators, a minimum leaf size of 2, and balanced class weights. The other seven classifiers used their model-specific defaults without class weighting. Thus, the nine-model comparison reflects both algorithm choice and model-specific handling of class imbalance rather than a uniform imbalance strategy. The radial-basis-function SVM used $C = 2$ with probability estimation, while the *k*-nearest-neighbor classifier used five distance-weighted neighbors. AdaBoost used 200 estimators and a learning rate of 0.05. Gradient boosting used 200 estimators, learning rate 0.05, and maximum depth 2. XGBoost used 200 estimators, maximum depth 3, a learning rate of 0.05, row and column subsampling of 0.9, and L2 regularization of 1. Random seed 12 was used wherever supported.

### 2.8 Cross-validation and model selection

The three missing-data strategies, three representations, and nine classifiers yielded 81 prespecified analyses. Evaluation in the development sample used five-fold repeated stratified cross-validation with three repeats and random seed 12, producing 15 validation-fold scores per combination. Row assignments for every repeat and fold were retained separately for the full development sample and its complete-case subset. Complete preprocessing pipelines, rather than previously transformed matrices, were submitted to cross-validation. The comparison table therefore contained training cross-validation metrics only.

Models were ranked by mean ROC area, followed by mean F1 score and mean accuracy, with each metric rounded to 12 decimal places. Configurations with identical rounded values were treated as tied under the stated ranking rule and reported as jointly top-ranked, together with the number of tied configurations. No tied configuration was described as uniquely best. An exact tie is a computational statement, not a claim of practical equivalence; conversely, configurations whose rounded values differed were ranked, but a difference smaller than the fold-to-fold variation was not treated as evidence that one configuration is superior (Section 4.1). When a figure or a downstream diagnostic required a single pipeline, one member of the tied set was selected according to a documented lexicographic ordering of scenario, imputation, and model name and designated the display representative. This designation reflects a technical choice rather than statistical superiority. Complete-case and imputed analyses were ranked and interpreted separately because they represented different analytical populations. No single pipeline was designated as primary. Among the 54 retained-sample candidates, the top-ranked configuration was PCA with median/mode imputation and Random Forest; the selected-original-feature representative is reported in greater detail because it retains named clinical fields, not because it ranked higher.

For each metric, the mean and sample standard deviation were calculated across the 15 validation folds. The reported fold-score range corresponds to the empirical 2.5th–97.5th percentiles of these scores. Because validation folds overlap across repeats, this range describes score stability rather than confidence in performance in an independent population [8,9].

### 2.9 Holdout evaluation and bootstrap analysis

Once both cross-validation rankings had been frozen, we fitted a prespecified set of finalists once to the corresponding complete development branch and evaluated them once on the matching holdout branch. This set included the display representative for each feature representation among strategies retaining all records; the complete-case display representative, the display representative among non-PCA tree models retaining all records, used for SHAP [15]; and the median/mode XGBoost pipelines with all original features and with PCA, included for the comparison with the published study. No other pipeline was evaluated on the holdout, and holdout results played no role in model ranking. Because the all-original-feature representative and the non-PCA tree representative were the same configuration, six unique pipelines were evaluated.

The reported measures were accuracy, positive predictive value, sensitivity, specificity, F1 score, ROC area, precision-recall area, and Brier score. Precision-recall area was implemented as average precision, and specificity was defined as TN/(TN + FP).

Holdout uncertainty was summarized using 1,000 stratified bootstrap resamples that preserved the observed numbers of CKD and non-CKD records. Percentile intervals were derived from predictions generated by the already fitted model; the model was not retrained within bootstrap samples. The conditional stratified-bootstrap percentile intervals, therefore, condition on the selected model, fixed partition, and observed class counts. They reflect sampling variation among the observed holdout records but do not capture uncertainty arising from model development or selection. The observed holdout metric, not the bootstrap mean, was used as the point estimate. An interval that collapses to a single value after perfect classification should not be interpreted as an absence of uncertainty. We therefore also report exact (Clopper–Pearson) 95% binomial intervals for holdout accuracy, computed from the confusion counts; they remain informative when every record is classified correctly but, like the bootstrap intervals, treat the fitted model as fixed.

### 2.10 Calibration

Calibration was assessed for the cross-validation-selected display representative of each feature representation among strategies that retained all holdout observations. Two binning schemes were used. Calibration curves were based on ten quantile bins of predicted probabilities. Because many probabilities were tied near 0 or 1, fewer than ten bins could be occupied; the figure therefore reports the number and size of the occupied bins together with the individual predicted probabilities. Bins were not connected by lines because no calibration was observed between them. The Brier score was calculated as the mean squared difference between predicted CKD probability and the binary outcome, with lower values indicating better probability accuracy [16]. A ten-bin expected calibration error (ECE) [17] was computed separately with equal-width bins on [0, 1] ECE was defined as the weighted mean absolute difference between observed and predicted risk across bins, with empty bins assigned zero weight. Given the small holdout size, ECE was treated as descriptive.

### 2.11 Ablation, ordinal-encoding sensitivity, and SHAP analyses

Leave-one-feature-out ablation was conducted for the all-original-feature representation using its own cross-validation display representative among analyses that retained all records. The model family and missing-data strategy were frozen from the selection manifest before any holdout evaluation. Each field was removed in turn, and the complete pipeline was reevaluated on the identical 15 persisted validation folds. Paired differences in accuracy relative to the all-feature pipeline were computed fold by fold and summarized by their mean and their 2.5th–97.5th percentiles across folds; because folds overlap, these percentiles are descriptive and tend to be narrow. All 24 fields are reported without truncation (Supplementary Figure S1). The same configuration (all original fields, k-nearest-neighbor imputation, Extra Trees) was also the non-PCA tree representative used for SHAP; ablation refits it within folds, whereas SHAP interprets a single fit on the full development sample.

For the ordinal-encoding sensitivity analysis, the all-original-feature display representative (the ablation pipeline: all original fields, k-nearest-neighbor imputation, Extra Trees) was refitted on the same 15 folds. Following within-fold mode imputation, specific gravity, albumin, and sugar were coded as ordered integers according to their prespecified clinical levels (1.005–1.025 and 0–5); all other categorical fields remained nominal. Integer coding assumes equal spacing in linear and distance-based models and imposes an ordering constraint in tree models. Paired fold differences from the nominal encoding were summarized as described above and did not influence model selection.

SHAP analysis was applied to the cross-validation-selected non-PCA tree model among approaches retaining all records [18]. The model was fitted on the development sample and interpreted using the holdout predictors [15]. Because one-hot encoding divides each categorical field into several model inputs, while the two levels of a binary field are exact complements, importance was summarized at the level of the original clinical variable. For each observation, the signed SHAP contributions of all inputs derived from the same variable were summed, and the mean absolute grouped contribution was reported. SHAP values are expressed on the predicted-probability scale. Values for transformed columns were retained as supplemental detail. SHAP magnitude was not treated as a causal effect or as evidence that a measurement is clinically necessary. Under *k*-nearest-neighbor imputation, a raw variable can also influence the imputed values of other variables.

### 2.12 Software environment

Analyses were run in Python 3.9.6 using NumPy 2.0.2, pandas 2.3.3, SciPy 1.13.1, scikit-learn 1.6.1 [19], XGBoost 2.1.4, SHAP 0.49.1, Matplotlib 3.9.4, Seaborn, and OpenPyXL. Exact versions of every package are recorded in the run manifest and in a lock file distributed with the code. Random seed 12 was used for all stochastic procedures unless otherwise specified. The released scripts regenerate the main-analysis figures, tables, and manuscript values from persisted machine-readable outputs through an evidence ledger that records the source file of every value; the values of the post hoc analyses (Section 2.14) were taken from the same persisted tables. The post hoc sensitivity analyses of Section 2.14 were run in the same environment.

### 2.13 Seed-sensitivity analysis

Because the development–holdout split, cross-validation folds, bootstrap resamples, and stochastic classifiers all depend on a single random seed, the main analysis (seed 12) was repeated from scratch with three additional prespecified seeds (42, 39, and 55). For each replication, the stratified 70:30 partition was redrawn, the population-specific repeated cross-validation registries were regenerated, all 81 combinations were re-evaluated, and the tie-aware selection rule was reapplied. Each frozen finalist was then fitted once to the corresponding development sample and evaluated once on the matching holdout sample. No bootstrap intervals were computed in the replications. The replications began only after the main selection manifest had been frozen, could not influence it, and are reported as descriptive evidence of the stability of the selection rule rather than as additional validation; the holdout samples of different seeds overlap and are not independent.

### 2.14 Post hoc sensitivity analyses

Seven additional sensitivity analyses were conducted after the selection manifest, the finalists, and the holdout evaluation had been frozen. None altered the main analysis. All used the persisted development-sample folds of Section 2.8 (the nested inner folds and the all-record grids were generated afresh), and, except for the descriptive missingness table, the selection-stability analysis (which pools the training folds of the three seed replications), the all-record naive-versus-controlled comparison, and the holdout demonstration described below, none used the holdout sample. They were archived as a separate, hash-verified stage of the analysis, and the controlled 81-pipeline grid recomputed within that stage reproduced the main-analysis fold scores exactly.

*Nested resampling.* To estimate the performance of the complete selection procedure rather than of any selected pipeline [20], the 15 persisted validation folds served as outer folds. Within each outer training fold, a new five-fold, three-repeat inner cross-validation was generated, every candidate pipeline of the relevant scope was re-evaluated, and the tie-aware selection rule of Section 2.8 was reapplied; only the pipeline selected in that fold was refitted to the outer training fold and evaluated once on the outer validation fold. Six scopes were examined: all 54 retained-sample candidates, each of the three representations, the non-PCA tree models, and the complete-case branch. Optimism was defined as the selected pipeline's inner cross-validation accuracy minus its outer accuracy, and outer performance was also compared, fold by fold, with the main-analysis display representative of the same scope. Because the 15 outer folds arise from three repeats of five-fold cross-validation, they overlap and are not independent; outer-fold ranges and paired differences are therefore descriptive and are not confidence intervals.

*Missingness pattern and multiple imputation.* Missingness of each field was tabulated by outcome class, and logistic regression and Extra Trees were fitted to the 24 binary missing-value indicators alone on the persisted folds. For the three retained-sample display representatives, numeric predictors were then imputed 20 times within each training fold by chained equations (Bayesian ridge regression with posterior sampling) [21], while categorical predictors kept the training-fold mode. Each imputed training set was fitted separately, validation-fold probabilities were averaged across the 20 imputations, and the pooled scores were compared with single imputation on identical folds.

*Feature-selection stability.* The eight-field supervised selector was refitted in every training fold of the main analysis and of the three seed replications (60 selection sets per missing-data strategy). Selection frequency of each field, mean pairwise Jaccard similarity, and the Nogueira stability index with a 2,000-resample bootstrap percentile interval were computed [22].

*Removal of diagnostic predictors.* The 24 fields were grouped by their clinical relation to CKD (Supplementary Table S6): renal-function markers, urinary damage markers, hematologic fields, electrolytes, metabolic and comorbidity fields, symptoms and signs, and demographic and vital-sign fields. Groups were removed singly and cumulatively, three subsets were retained alone, and the top one to eight fields of the grouped SHAP ranking were removed cumulatively. Each reduced pipeline was re-evaluated on the same 15 folds for the all-original and selected-feature Extra Trees representatives, with supervised selection repeated within folds, and paired fold differences from the full pipeline were summarized. Because the dataset carries no timestamps, whether each measurement preceded the diagnosis could not be verified empirically; the grouping records the expected temporal relation instead.

*Uniform class weighting.* All 81 pipelines were re-evaluated under two uniform policies: no weighting for any classifier, and balanced class weights passed as training-fold sample weights to every classifier that accepts them (all except *k*-nearest neighbors). Paired fold differences between the policies were summarized by algorithm, the selection rule was reapplied under each policy, and pipelines whose policy matched the main analysis were verified to reproduce its fold scores exactly.

*Naive versus controlled analysis.* The same 81 pipelines were re-evaluated on identical folds after imputation, scaling, feature selection, and PCA had been fitted once on all rows before cross-validation—the leakage that the main design prevents—both on the development sample ($n = 280$) and on all 400 records. Rankings were compared by Spearman correlation of mean accuracies quantized to 12 decimal places, as in the selection rule, so that exactly tied pipelines received tied ranks. As a further demonstration, every controlled pipeline was evaluated once on the holdout sample and the best holdout accuracy among the 54 retained-sample pipelines was recorded (the 27 complete-case pipelines were scored on their 46-record holdout subset), mimicking selection on the test sample.

Table 1. Experimental comparisons and safeguards against data leakage

| Component | Specification | Information boundary |
|---|---|---|
| Development–holdout split | Stratified 70:30 partition; seed 12; row assignments persisted | Completed before learned preprocessing; holdout is a locked historical internal sample |
| Missing data | Median/mode; standardization followed by numeric five-neighbor imputation with categorical mode; complete-case deletion | Imputation and scaling fitted within training folds; deletion applied separately to each partition |
| Feature representation | All 24 fields; level-weighted PCA of the expanded encoded matrix to ≥95% variance; mutual-information selection of eight named fields | Encoding, scaling, PCA, and selection fitted within training folds |
| Classifiers | Nine fixed algorithms; 81 total combinations | Holdout outcomes excluded from fitting and ranking |
| Internal validation | Five folds × three repeats on persisted folds; ranking by ROC area, F1, then accuracy with exact ties reported | Development sample only; no holdout column in the comparison table |
| Holdout evaluation | Frozen finalists evaluated once; discrimination, calibration, Brier score, and 1,000 stratified bootstrap resamples | Conditional evaluation of fixed fitted models after all rankings were frozen |

### 2.15 Use of artificial-intelligence tools

Two large-language-model assistants were used under the direction of the authors through their macOS desktop applications: OpenAI Codex (GPT-5.5), from July to August 2026, and Claude (Anthropic; Claude Fable 5 and Claude Fable 5.1), from August to September 2026. They were used for two purposes. First, Codex was used to modify and extend the Python analysis code written by the authors (bug fixes, additional functions and tests, and refactoring). Second, Claude was used to audit the code for data leakage and to check every value reported in this article against its source file in the persisted outputs; the corrections arising from that audit—fixes to the sensitivity-analysis code, removal of unused files, a figure-rendering script, and tracked-change edits to the reported values—were implemented with Claude and reviewed by the authors. The study design, the prespecified selection rule, the leakage controls, and the interpretation of the results were decided by the authors. All code was executed by the authors on their own hardware; every reported value derives from machine-readable outputs that are hash-verified within the released run (run_2026-09-28_seed12), and the code was validated by an automated test suite (359 unit tests and 25 release tests at the time of release) and by the authors' review. Neither assistant generated data or produced any result value: every value was obtained by executing the code, and the assistants' edits to the manuscript transcribed values from the persisted outputs. The authors take full responsibility for the content of this article; the AI tools are not authors.

## 3. Results

### 3.1 Dataset composition and missingness

The analytical dataset comprised 250 CKD records (62.5%) and 150 non-CKD records (37.5%). Missingness differed considerably across predictors (Table 2). Red blood-cell morphology (152/400, 38.0%) had the highest proportion of missing observations, followed by red blood cell count (131/400, 32.8%), white blood cell count

(106/400, 26.5%), potassium (88/400, 22.0%), and sodium (87/400, 21.8%). Complete-case deletion retained 158 of 400 records (39.5%): 115 non-CKD records (72.8% of the retained sample) and 43 CKD records (17.2% of all CKD records). Applied separately to the fixed partitions, deletion retained 112 development records (81 non-CKD and 31 CKD) and 46 holdout records, compared with 280 and 120, respectively, under imputation. Thus, complete-case deletion reduced the sample and reversed the majority class as well. The proportion of CKD fell from 62.5% in the full development sample to 27.7% in the complete-case development sample, making accuracy, precision, precision-recall area, and Brier score incomparable across the two populations.

Table 2. Predictor groups and missingness in the analytical dataset ($N$ = 400)

| **Predictor group** | **Variables** | **Missing observations, *n* (%)** |
|---|---|---|
| Demographic and vital signs | Age; blood pressure | 9 (2.3%); 12 (3.0%) |
| Metabolic and renal laboratory measures | Random blood glucose; blood urea; serum creatinine; sodium; potassium | 44 (11.0%); 19 (4.8%); 17 (4.3%); 87 (21.8%); 88 (22.0%) |
| Hematologic measures | Hemoglobin; packed cell volume; white blood cell count; red blood cell count | 52 (13.0%); 71 (17.8%); 106 (26.5%); 131 (32.8%) |
| Urinalysis | Specific gravity; albumin; sugar; red blood cells; pus cell; pus cell clumps; bacteria | 47 (11.8%); 46 (11.5%); 49 (12.3%); 152 (38.0%); 65 (16.3%); 4 (1.0%); 4 (1.0%) |
| Clinical history and examination | Hypertension; diabetes mellitus; coronary artery disease; appetite; pedal edema; anemia | 2 (0.5%); 2 (0.5%); 2 (0.5%); 1 (0.3%); 1 (0.3%); 1 (0.3%) |

## 3.2 Reproduction of the published comparison

The source study reported XGBoost accuracies of approximately 0.983 with the original data and 0.9916 with data described as PCA-based [3]. In the leakage-controlled reproduction, which used median/mode imputation and a single five-fold stratified cross-validation of the development sample, mean XGBoost accuracy was 0.968 for the original representation and 0.986 for PCA. Both prespecified pipelines achieved an accuracy of 1.000 on the locked holdout. PCA therefore gave a higher development cross-validation accuracy but not a higher holdout accuracy. Under the repeated five-fold cross-validation, the same XGBoost pipelines reached 0.9738 with the original representation and 0.9893 with PCA (mean paired fold difference, 0.0155; 2.5th–97.5th fold percentiles, −0.0116 to +0.0357), and the difference was 0.0214, 0.0012, and 0.0179 under seeds 42, 39, and 55; on the replication holdouts, PCA was higher under two seeds and lower under one. The direction of the published PCA advantage for XGBoost was therefore reproduced. PCA likewise raised the accuracy of the other boosting, bagging, decision-tree, and nearest-neighbor classifiers (by about 0.01–0.04 on average across seeds), changed that of Extra Trees and the SVM by less than 0.01, and lowered that of logistic regression. In the full repeated-cross-validation comparison (Section 3.5), however, the PCA display representative exceeded the best original-feature representative by only 0.0036 in accuracy (computed before rounding), less than the fold-to-fold standard deviation of either pipeline. The high overall classification performance was broadly reproduced, but relative to the best original-feature pipelines, the reported advantage of PCA emerged only as a marginal, seed-dependent difference (Section 3.8) that narrowed to 0.0024 and lay within fold-to-fold variation when the selection procedure was nested within cross-validation (Section 3.9). Direct numerical comparison remains limited because of the PCA-labelled presentation in the source article; the published PCA-labelled representation retained named clinical variables, whereas the current PCA representation consisted of standardized linear components of the expanded encoded matrix.

## 3.3 Repeated cross-validation

Across the 81 combinations, 8 configurations tied exactly at the top of the prespecified cross-validation ranking, each with an ROC area, F1 score, and accuracy of 1.0000. All 8 used complete-case deletion, representing 8 of the 27 complete-case configurations. This ceiling likely reflects the small and readily separable complete-case development sample of 112 records rather than any distinctive property of a particular classifier; ranks within the tied set are therefore uninformative. The display representative - Extra Trees with all original features and complete-case deletion - is simply one member of this set. Because deletion removed 60.5% of the cohort, reversed the class balance (its holdout subset contained 34 non-CKD and 12 CKD records), and produced perfect validation scores for many classifiers, the complete-case branch was treated as a sensitivity analysis rather than the primary result.

Among approaches that retained all records, a single configuration ranked first for the selected-original-feature representation: selected original features with numeric $k$-nearest-neighbor imputation and Extra Trees. The all-original and PCA representations likewise each had one top-ranked configuration. The selected-feature representative achieved a repeated-cross-validation accuracy of 0.9869 ± 0.0218, with fold scores ranging from 0.9348 to 1.0000. Mean sensitivity was 0.9829, specificity 0.9937, F1 score 0.9892, ROC area 1.0000, and precision-recall area 1.0000. Extra Trees using all original features and numeric $k$-nearest-neighbor imputation achieved an accuracy of 0.9905 ± 0.0133, whereas Random Forest using PCA and median/mode imputation achieved 0.9940 ± 0.0087 (Table 3). The complete-case sensitivity analysis is reported separately in Table 4. Figure 2 compares repeated-cross-validation and holdout accuracy across the four display representatives and presents the cumulative variance explained by PCA.

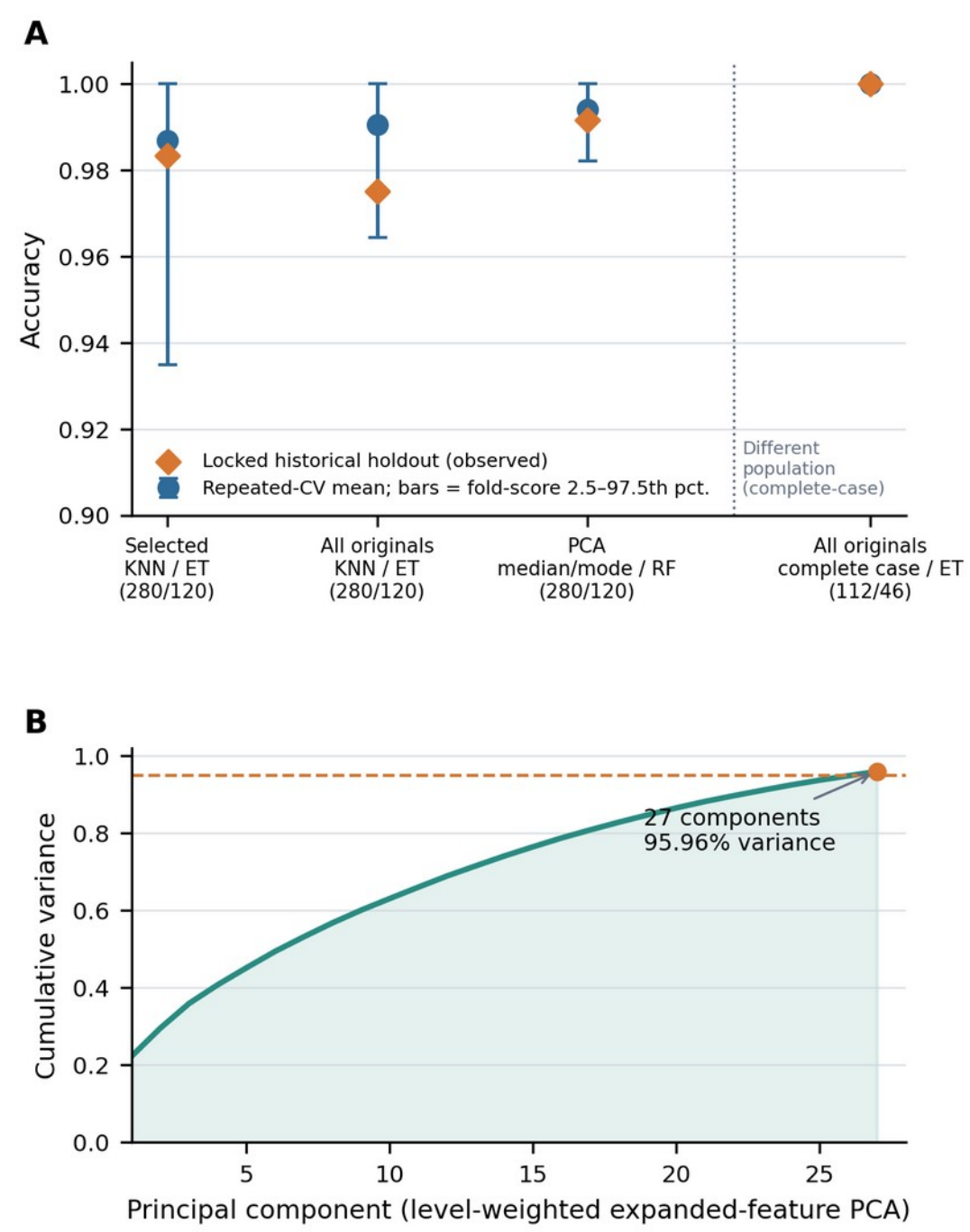

*Figure 2. Comparative performance and PCA dimensionality. (A) Mean repeated-cross-validation accuracy with error bars spanning the 2.5th–97.5th percentiles of the 15 validation-fold scores (bounded by 0 and 1) and observed accuracy on the locked historical holdout for the four display representatives. The complete-case result, separated by the dotted line, refers to a different analytical population (112 development and 46 holdout records) and is not population-equivalent to the imputed analyses. (B) Cumulative explained variance of the principal components fitted to the full development sample; 27 components retained at least 95% of total variance.*

Table 3. Principal repeated-cross-validation and locked-holdout results (analyses retaining all records)

| **Pipeline and sample size (development/holdout)** | **Repeated cross-validation** | **Locked historical holdout** |
|---|---|---|
| Selected originals; KNN; Extra Trees (280/120) | Accuracy 0.9869 ± 0.0218 (fold-score range, 0.9348–1.0000); sensitivity/specificity 0.9829/0.9937; ROC area 1.0000 | Accuracy 0.9833 (conditional bootstrap interval, 0.9583–1.0000; exact binomial interval, 0.9411–0.9980); ROC area 1.0000; Brier score 0.0142 |
| All originals; KNN; Extra Trees (280/120) | Accuracy 0.9905 ± 0.0133 (fold-score range, 0.9643–1.0000); sensitivity/specificity 0.9848/1.0000; ROC area 1.0000 | Accuracy 0.9750 (conditional bootstrap interval, 0.9500–1.0000; exact binomial interval, 0.9287–0.9948); ROC area 1.0000; Brier score 0.0149 |
| PCA ≥95%; median/mode; Random Forest (280/120) | Accuracy 0.9940 ± 0.0087 (fold-score range, 0.9821–1.0000); sensitivity/specificity 0.9905/1.0000; ROC area 1.0000 | Accuracy 0.9917 (conditional bootstrap interval, 0.9750–1.0000; exact binomial interval, 0.9544–0.9998); ROC area 1.0000; Brier score 0.0053 |

*KNN, k-nearest-neighbor imputation; PCA, principal component analysis; SVM, support-vector machine. Each row is the display representative of its representation among analyses retaining all records. Fold-score ranges are the 2.5th–97.5th percentiles of 15 overlapping validation folds and do not estimate uncertainty in an external population. Bootstrap intervals are conditional on fixed fitted models and on the observed class counts. Exact binomial intervals are Clopper–Pearson 95% intervals for the number of correctly classified holdout records.*

Table 4. Complete-case sensitivity analysis in a different analytical population

| **Pipeline and sample size (development/holdout)** | **Repeated cross-validation** | **Locked historical holdout** |
|---|---|---|
| All originals; complete case; Extra Trees (112/46); development 81 non-CKD and 31 CKD; holdout 34 non-CKD and 12 CKD | Accuracy 1.0000 ± 0.0000 (fold-score range, 1.0000–1.0000); sensitivity/specificity 1.0000/1.0000; ROC area 1.0000 | Accuracy 1.0000 (conditional bootstrap interval, 1.0000–1.0000; exact binomial interval, 0.9229–1.0000); ROC area 1.0000; Brier score 0.0030 |

*The complete-case representative is one of 8 exactly tied configurations and was evaluated in a subset in which non-CKD is the majority class. Accuracy, precision, precision-recall area, and Brier score are prevalence dependent and are therefore not comparable with Table 3; ROC area is the only metric intended for cross-population comparison. The exact binomial interval is the Clopper–Pearson 95% interval for the number of correctly classified holdout records.*

## 3.4 Holdout discrimination and bootstrap intervals

The selected-feature Extra Trees representative correctly classified 118 of 120 holdout records (exact 95% interval, 0.9411–0.9980): TN = 45, FP = 0, FN = 2, and TP = 73. Its precision-recall area was 1.0000.

The all-feature Extra Trees classified 117 of 120 records correctly (FN = 3, FP = 0; exact interval, 0.9287–0.9948), and the PCA Random Forest classified 119 of 120 records correctly (FN = 1, FP = 0; exact interval, 0.9544–0.9998). The exact intervals of the five finalists evaluated on the full holdout overlapped, including the interval of 0.9697–1.0000 for the two XGBoost pipelines, which each classified all 120 records correctly; a holdout of this size cannot distinguish among these finalists. The complete-case Extra Trees classified 46 of 46 records in its smaller holdout subset correctly (FN = 0, FP = 0; exact interval, 0.9229–1.0000). Because it classified every record correctly, its fixed-prediction bootstrap interval collapses to a single value; this reflects perfect classification of the observed records under conditional resampling, not an absence of uncertainty in model development or transport to new settings. Figure 3 presents the conditional bootstrap intervals for the four display representatives.

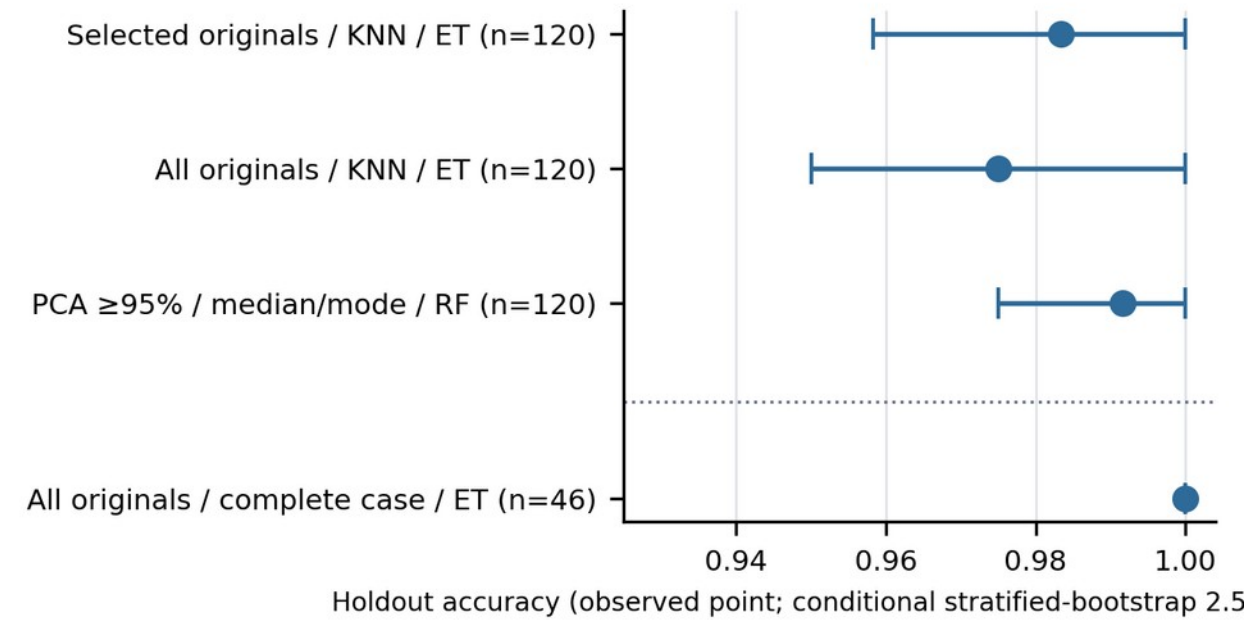

*Figure 3. Conditional stratified-bootstrap intervals for holdout accuracy. Points indicate observed accuracy and bars indicate the 2.5th–97.5th percentiles from 1,000 class-stratified resamples of the fixed predictions. Models were not retrained during the bootstrap procedure, and the complete-case pipeline (below the dotted line) was evaluated in a different, smaller holdout population. Exact binomial intervals for these accuracies are given in Tables 3 and 4.*

## 3.5 Feature representation and missing-data sensitivity

Among analyses of the methods retaining all records, the cross-validation accuracies of the three representation-specific display representatives were closely grouped: 0.9869 for selected original fields, 0.9905 for all original fields, and 0.9940 for PCA. Fitted to the full development sample, the PCA model retained 27 components that explained 95.96% of the variance across 47 standardized expanded columns. PCA therefore had a slightly higher mean cross-validation accuracy than the best original-feature representation in the main analysis but used more transformed dimensions than the eight-field supervised selector. The number of components is not directly comparable with the number of original fields because one-hot encoding expands categorical predictors.

The selected-feature and all-original-feature representatives used numeric k-nearest-neighbor imputation, whereas the PCA representative used median/mode imputation. Differences between imputation strategies were small relative to fold-to-fold variation. Complete-case analyses often produced perfect validation scores, but they evaluated a much smaller and differently composed subset. Their apparent performance does not show that deleting incomplete records is preferable to imputation.

## 3.6 Calibration

All three representative retained-sample models had low Brier scores on the holdout sample. The lowest Brier score, 0.0053, was observed for the PCA Random Forest pipeline; the all-original-feature Extra Trees had a Brier score of 0.0149 and an ECE of 0.0500; the PCA Random Forest, 0.0053 and 0.0325; and the selected-feature Extra Trees, 0.0142 and 0.0466. Most predicted probabilities clustered near 0 or 1, and tied values collapsed adjacent quantile bins. The ten requested bins yielded 8, 10, and 9 occupied bins for the three models, respectively, leaving little calibration information across wide intermediate probability ranges (Figure 4). Deviations from the ideal line occurred in sparsely populated bins, while equal-width ECE depended on the placement of the few intermediate probabilities. With only 120 holdout observations, small differences in Brier score or ECE should not determine model choice.

## 3.7 Ablation and SHAP findings

The all-feature pipeline used for ablation - Extra Trees with numeric *k*-nearest-neighbor imputation and the display representative of the all-original-feature representation - achieved a repeated-cross-validation accuracy of 0.9905 before any field was removed. Removing specific gravity produced the largest mean paired decrease in accuracy (−0.0238; fold-difference range, −0.0768 to 0.0000). The range included zero for all 24 fields and remained entirely below zero for none. Consequently, the effect of removing any single field could not be distinguished from fold-to-fold variation. Supplementary Figure S1 presents the complete set of paired differences without truncation.

In the all-original-feature / KNN / Extra Trees model, hypertension had the largest grouped mean absolute SHAP contribution (0.0987), followed closely by albumin (0.0907) and then diabetes (Figure 5). The first two values were similar and should not be treated as meaningfully ranked. At the level of individual transformed inputs, albumin = 0 made the largest contribution (0.0840). This level-wise ordering differs from the variable-level ranking because categorical fields are divided across several inputs and the two levels of a binary field are complementary. The leading variables also differed between the grouped SHAP and ablation rankings. These analyses capture different aspects of model behavior - attribution within a fitted model and performance after retraining an all-feature model without one field - and neither supports causal interpretation.

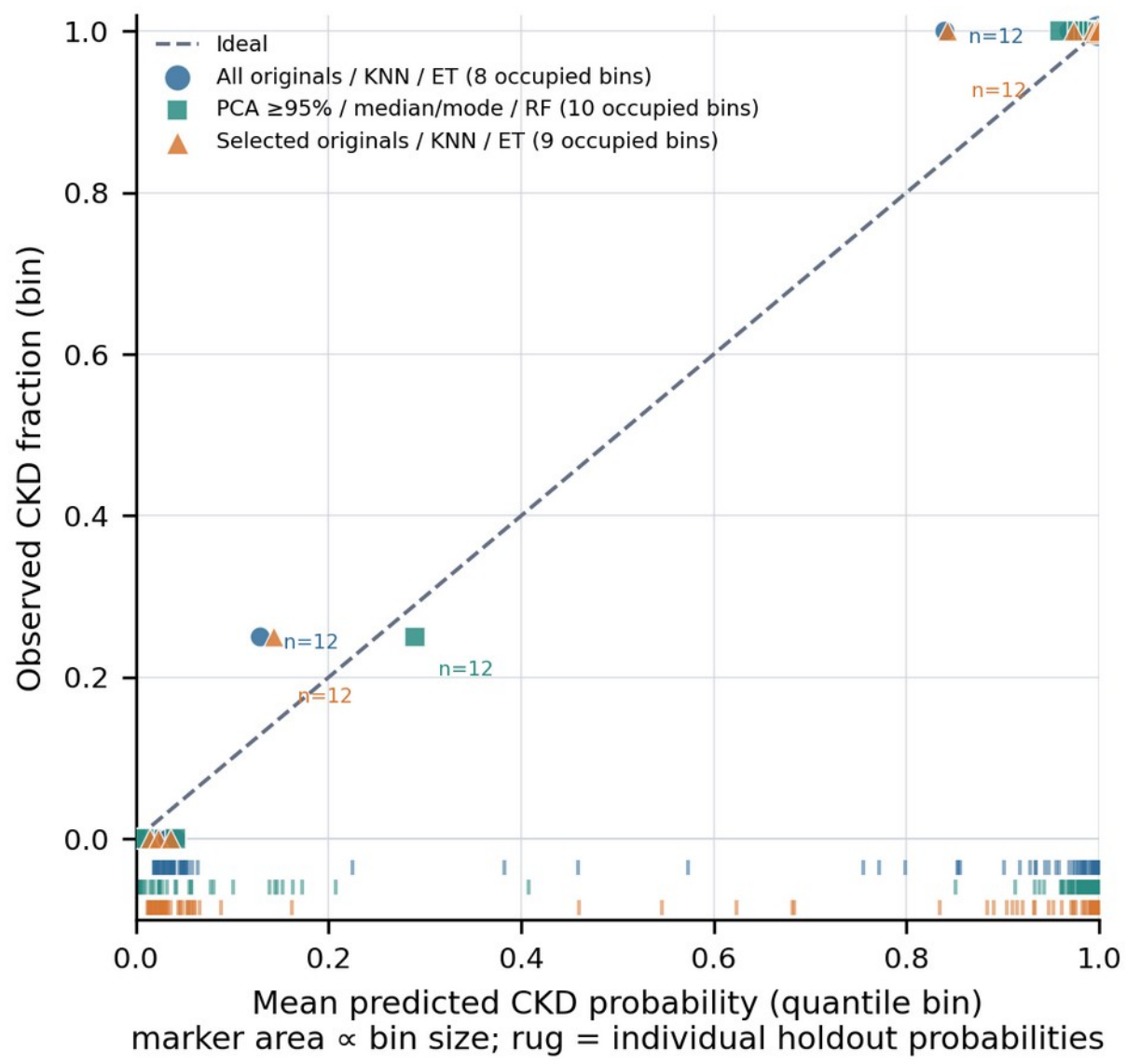

*Figure 4. Holdout calibration of the representative analyses retaining all records. Markers show quantile-based bins with the number of holdout records in each bin; bins are not connected because no calibration is observed between them, and the rug below the axis shows the individual predicted probabilities. The dashed diagonal denotes ideal calibration. Lower Brier scores indicate better probability accuracy.*

## 3.8 Seed sensitivity

Across the three additional seeds (42, 39, and 55) and the five selection scopes - 15 seed-scope pairs in total - the display representative of the main analysis remained jointly top-ranked in 6 pairs and was the identical display representative in 6 pairs; its worst dense rank in any replication was 6 (Supplementary Table S1). For the selected-feature representation highlighted in the main text, the main-analysis configuration remained jointly top-ranked under two of the additional seeds, and the display representatives of that scope reached repeated-cross-validation accuracies of 0.9810–0.9952 and holdout accuracies of 0.9750–0.9833 across the replications. Among all retained-sample scopes and replications, the lowest holdout accuracy of any display representative was 0.9750 and the lowest holdout ROC area was 0.9985. When the main representative fell below the top rank, Supplementary Table S1 reports its dense rank and the cross-validation and holdout metrics of the displayed alternative.

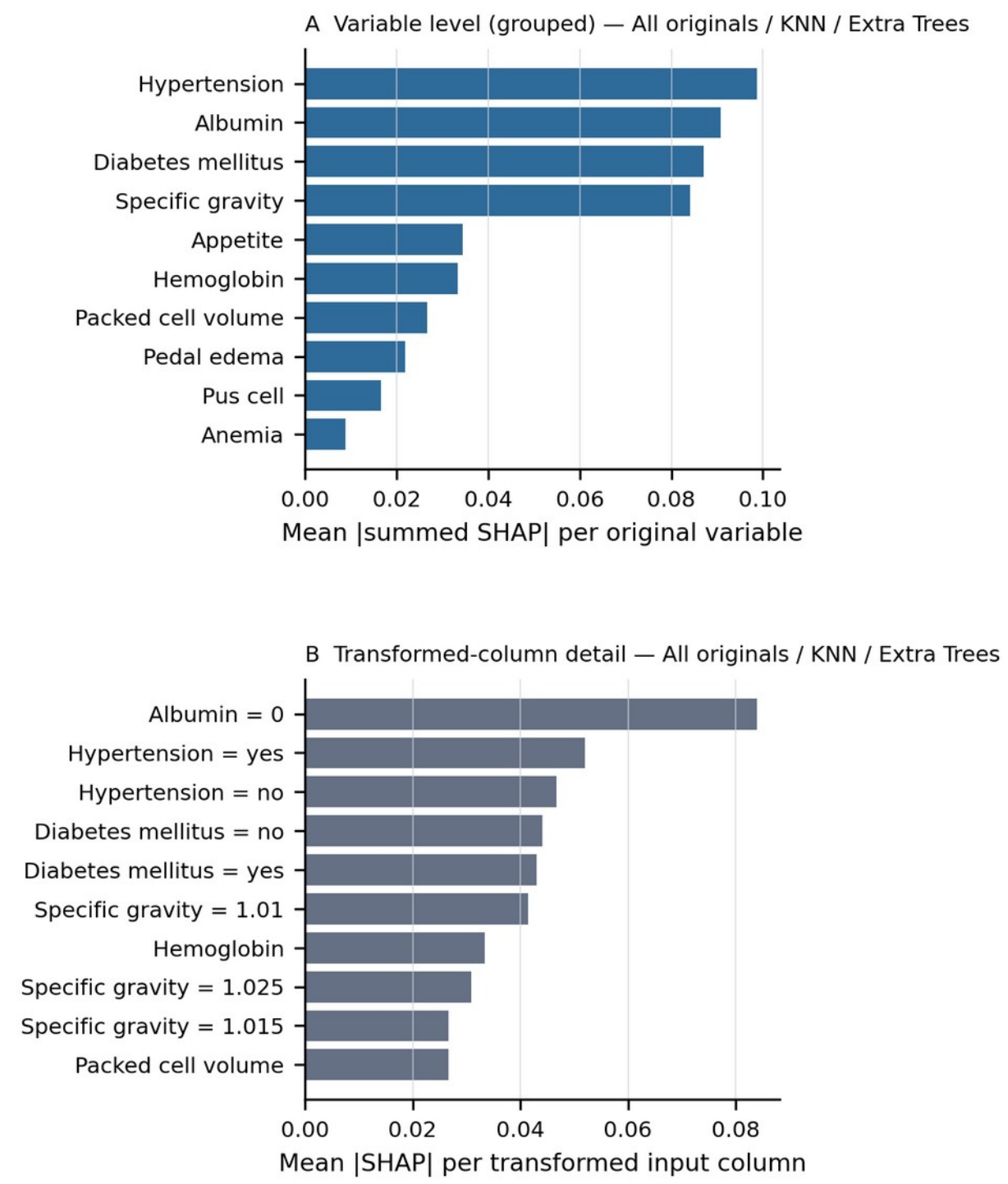

*Figure 5. Model interpretation for the all originals / KNN / Extra Trees model, fitted on the development sample and explained on the holdout predictors. (A) Mean absolute grouped SHAP contribution per original clinical variable, obtained by summing the signed contributions of all transformed inputs derived from each variable within every observation. (B) Mean absolute SHAP value per transformed input column (one-hot levels shown separately) as supplemental detail. The ablation analysis (Supplementary Figure S1) refits the same configuration within folds.*

## 3.9 Post hoc sensitivity analyses

Nested resampling of the complete selection procedure gave outer accuracies of 0.9845–0.9881 across the five retained-sample scopes, compared with 0.9869–0.9940 for the main display representatives on the same folds (Table 5; Supplementary Figure S2). The optimism of the selected pipeline's inner cross-validation estimate ranged from 0.0010 (selected-feature scope) to 0.0092 (all 54 retained-sample candidates). When all retained-sample candidates were eligible, the procedure selected 9 distinct pipelines across the 15 outer folds; the main representative (PCA, median/mode, Random Forest) was selected in 2 folds, and its inner dense rank was as low as 12 in one fold. Outer ROC areas were 0.998–1.000. The nested outer accuracies of the PCA-specific (0.9869) and original-feature-specific (0.9845) procedures differed by 0.0024, and their outer-fold ranges overlapped almost entirely. Because the outer folds are the three repeats of the same five-fold partition, these ranges describe fold-to-fold variation and are not independent replicates.

Table 5. Nested resampling of the complete selection procedure on the development sample (15 outer folds)

| Selection scope (candidates) | Nested outer accuracy, mean ± SD (fold-score range) | Inner accuracy of selected pipeline; optimism | Main representative, same folds; nested − main (fold range) | Distinct pipelines selected; folds selecting main representative |
|---|---|---|---|---|
| All retained-sample candidates (54) | 0.9845 ± 0.0177 (0.9527–1.0000) | 0.9938; +0.0092 | 0.9940; −0.0095 (−0.0411 to +0.0116) | 9; 2/15 |
| All original fields (18) | 0.9845 ± 0.0177 (0.9527–1.0000) | 0.9915; +0.0070 | 0.9905; −0.0060 (−0.0295 to 0.0000) | 4; 1/15 |
| PCA ≥95% (18) | 0.9869 ± 0.0143 (0.9643–1.0000) | 0.9934; +0.0065 | 0.9940; −0.0071 (−0.0179 to 0.0000) | 6; 4/15 |
| Selected original fields (18) | 0.9845 ± 0.0222 (0.9348–1.0000) | 0.9855; +0.0010 | 0.9869; −0.0024 (−0.0179 to +0.0116) | 5; 6/15 |

| Selection scope (candidates) | Nested outer accuracy, mean ± SD (fold-score range) | Inner accuracy of selected pipeline; optimism | Main representative, same folds; nested − main (fold range) | Distinct pipelines selected; folds selecting main representative |
|---|---|---|---|---|
| Non-PCA tree models (20) | 0.9881 ± 0.0161 (0.9527–1.0000) | 0.9920; +0.0039 | 0.9905; −0.0024 (−0.0179 to 0.0000) | 3; 1/15 |
| Complete-case branch (27)† | 0.9970 ± 0.0117 (0.9705–1.0000) | 1.0000; +0.0030 | 1.0000; −0.0030 (−0.0295 to 0.0000) | 3; 3/15 |

*SD, standard deviation. Outer folds are the persisted 15 validation folds of the main analysis; within each outer training fold the five-fold, three-repeat inner cross-validation was regenerated and the tie-aware selection rule reapplied to the candidates of the scope. Fold-score ranges are the 2.5th–97.5th percentiles of 15 overlapping folds and do not estimate uncertainty in an external population. Optimism is the selected pipeline's inner cross-validation accuracy minus its outer accuracy. †Complete-case scope evaluated in a different population (n = 112) and shown for comparison only. Differences were computed before rounding.*

Missingness was strongly associated with CKD status (Supplementary Table S2). CKD records lacked a mean of 3.63 fields (median, 3) compared with 0.69 (median, 0) for non-CKD records; 17.2% of CKD records and 76.7% of non-CKD records were complete, and for the six fields with the most missing values the missingness rate was 8–11 times higher in CKD records. Logistic regression and Extra Trees fitted to the 24 missing-value indicators alone achieved cross-validation accuracies of 0.813 and 0.820, respectively, and ROC areas of 0.854 and 0.878, respectively. Multiple imputation changed the pooled accuracy of the retained-sample representatives by 0.0000 (all originals), −0.0012 (PCA), and −0.0060 (selected features) relative to single imputation, with 2.5th–97.5th fold percentiles of the paired differences of −0.0295 and +0.0116 and pooled ROC areas of 0.9998–1.0000 (Supplementary Table S3).

The eight-field selector was stable (Supplementary Table S4; Supplementary Figure S3). Across 60 training folds and four seeds, the Nogueira stability index was 0.902 (bootstrap interval, 0.873–0.936) under median/mode imputation, 0.875 (0.858–0.896) under *k*-nearest-neighbor imputation, and 0.869 (0.848–0.897) under complete-case deletion. Serum creatinine, hemoglobin, packed cell volume, and red blood cell count were selected in every fold under all three strategies, specific gravity in every imputed fold, albumin and hypertension in at least 95% of imputed folds, and sodium in 53–77%; no other field was selected in more than one third of imputed folds. The panel fitted to the full development sample had a mean Jaccard similarity of 0.93 (median/mode) and 0.89 (*k*-nearest neighbor) with the fold-level panels.

Removing any single group of predictors changed accuracy by less than 0.05, whereas removing the renal, urinary, and hematologic fields together, or retaining only a small subset of the fields, reduced it to 0.84–0.89 (Figure 6; Supplementary Table S5). For the all-original Extra Trees representative, removing serum creatinine and blood urea changed accuracy by −0.0012, and removing the six urinalysis markers by −0.0345; removing the renal-function, urinary, and hematologic fields together (11 fields retained) reduced accuracy to 0.8917 (ROC area, 0.9794), and the seven demographic, vital-sign, metabolic, and comorbidity fields alone yielded 0.8369 (ROC area, 0.9594). Serum creatinine with albumin alone gave 0.8750 (ROC area, 0.9519), and cumulative removal of the top eight grouped-SHAP fields lowered accuracy to 0.9071 (ROC area, 0.9861). Results for the selected-feature representative, with selection repeated within folds, were similar.

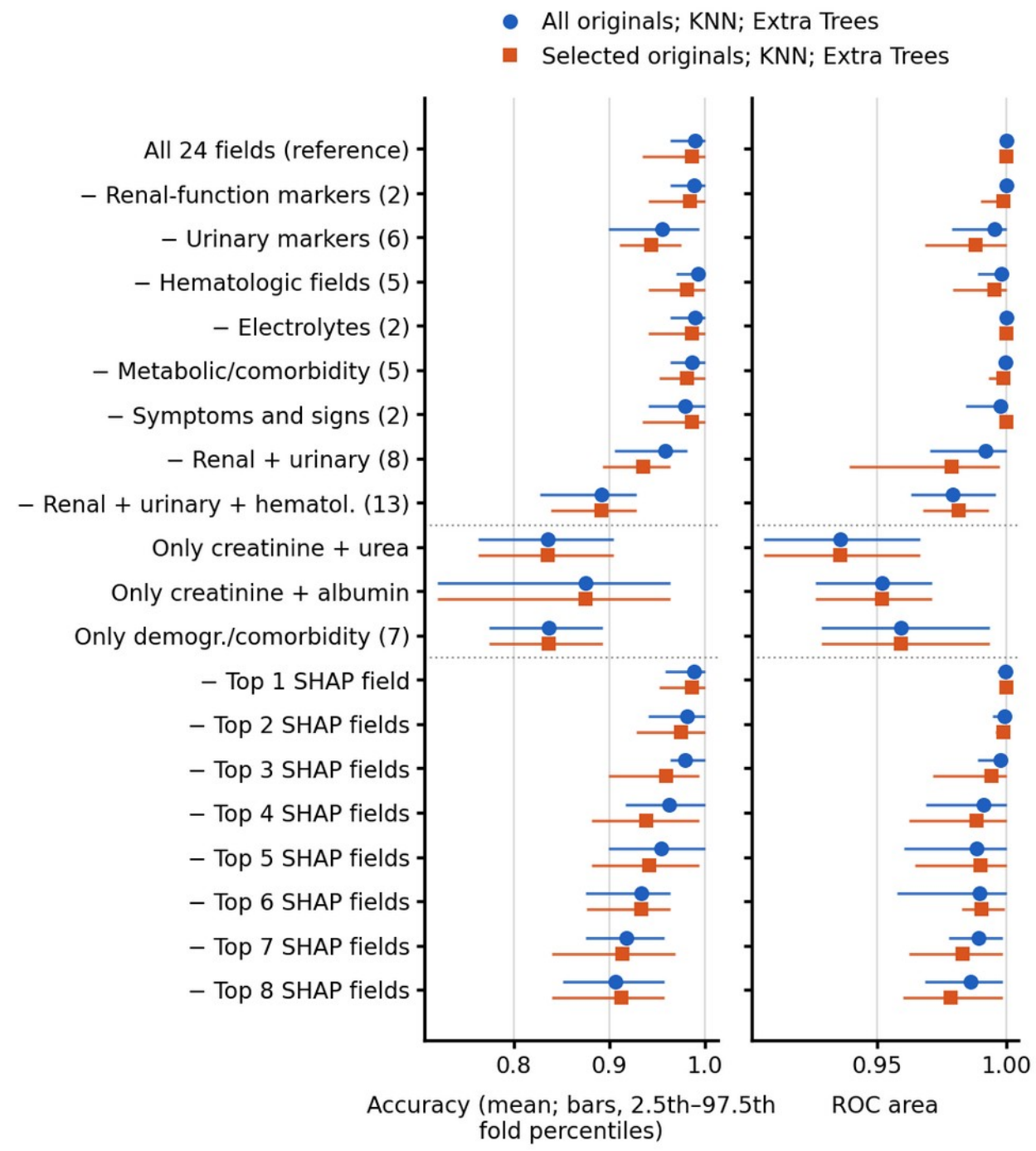

*Figure 6. Repeated-cross-validation performance after removal of diagnostic predictors on the development folds. Points show the mean over the 15 persisted folds and bars the 2.5th–97.5th fold percentiles for the all-original (circles) and selected-feature (squares) Extra Trees pipelines with numeric k-nearest-neighbor imputation; supervised selection was repeated within folds. Rows are clinical groups removed (numbers of fields in parentheses), subsets retained alone, and cumulative removal of the top-ranked grouped-SHAP fields. The holdout sample was not used.*

Applying a uniform class-weighting policy to all classifiers changed mean cross-validation accuracy by −0.0044 (AdaBoost) to +0.0044 (decision tree) per algorithm across the six retained-sample designs, and ROC area by at most 0.0031 (Supplementary Table S7). Under balanced weights for every classifier, reapplying the selection rule reproduced all six display representatives of the main analysis; with no weighting for any classifier, the all-original, selected-feature, and non-PCA tree scopes selected a median/mode Extra Trees pipeline instead, and the main representatives ranked second. Retained-sample accuracy rankings under either policy correlated with the main analysis at Spearman ρ ≥ 0.98.

Fitting imputation, scaling, feature selection, and PCA once on all development rows before cross-validation changed the mean accuracy of the 81 pipelines by between −0.0016 and +0.0029 within each representation–imputation design (range across individual pipelines, −0.0087 to +0.0143). The number of pipelines with a perfect ROC area rose from 8 to 11, the complete-case tie count from 8 to 9, the top-ranked retained-sample pipeline changed from PCA/median-mode/Random Forest to PCA/k-nearest-neighbor/Random Forest, and the rank correlation with the controlled ranking was 0.952 (Supplementary Table S8). Selecting among the 54 retained-sample controlled pipelines by holdout accuracy would have reported 1.0000 (8 pipelines tied; 23 of the 27 complete-case pipelines also reached 1.0000 on their 46-record subset), whereas the frozen finalists achieved 0.9750–1.0000.

## 4. Discussion

### 4.1 Principal findings

Very high internal discrimination on the UCI CKD dataset persisted when learned preprocessing was restricted to training folds and *k*-nearest-neighbor imputation was applied to standardized values. Among approaches retaining all observations, the representation-specific display representatives were nearly indistinguishable: mean repeated-cross-validation accuracies spanned only 0.0071, ROC areas reached or approached 1.0 across multiple representations, and the complete-case branch yielded numerous exact ties. With only 400 records, ranking methods on differences in the third decimal place would overstate the precision of model comparison; the present analysis therefore reports ties explicitly and treats the display representatives as technical examples. Two kinds of indistinguishability therefore appear in the results: exact computational ties, which are reported as such, and numerically different scores whose gaps lie within fold-to-fold variation, such as the 0.0071 spread among the retained-sample representatives against fold-to-fold standard deviations of 0.0087–0.0218; neither provides evidence that one pipeline is superior. Nested resampling of the entire selection procedure gave outer accuracies up to about 0.01 below the apparent cross-validation accuracy of the selected pipelines and selected different pipelines in different outer folds, so the display representatives are examples drawn from a performance plateau rather than demonstrably superior models.

Discrimination and probability accuracy do not necessarily produce the same ranking. In this analysis, however, both the lowest holdout Brier score and the highest repeated-cross-validation accuracy among retained-sample representatives belonged to the PCA Random Forest pipeline. If predicted probabilities are intended to guide clinical risk thresholds rather than binary classification alone, calibration requires independent evaluation and likely recalibration in the intended setting.

### 4.2 Comparison with the published study

The findings support the broad conclusion of Islam et al. that CKD labels in this dataset are highly separable using conventional machine-learning methods [3]. However, the advantage of PCA depended on the classifier. PCA raised the repeated-cross-validation accuracy of XGBoost (Section 3.2), consistent in direction with the published comparison, but relative to the best original-feature pipelines its advantage was marginal and seed dependent. Among analyses retaining all records, the PCA representative had the strictly highest repeated-cross-validation accuracy under three of the four seeds but the strictly highest holdout accuracy under one; all differences remained within fold-to-fold variation. When the PCA-specific and original-feature-specific selection procedures were nested within cross-validation, their outer accuracies were 0.9869 and 0.9845, respectively; the data therefore do not support a claim that PCA improves the best achievable prediction in this dataset. The standardized PCA representation comprised 27 orthogonal components, while XGBoost achieved the same holdout accuracy with PCA and the original representation (1.000). Differences may arise from data partitioning, missing-data handling, model configuration, or the distinction between mathematical PCA and the named-variable reduction presented in the source report.

The development–holdout partition and preprocessing rules were retained regardless of agreement with the published estimate of 99.16%. Reproducibility was defined by transparent implementation of the reported design and explicit documentation of discrepancies, not by numerical tuning toward the earlier value.

### 4.3 Leakage control and evaluation design

The principal safeguard against leakage was structural: every transformation that estimated a statistic or used the outcome was fitted within the model during each cross-validation iteration. Median and mode values, scaling parameters, neighbor distances, category vocabularies, principal axes, and mutual-information scores were therefore derived exclusively from the relevant training fold. Holdout observations were excluded from fitting and ranking, comparison tables contained no holdout column, and, in the main analysis, only the frozen finalist set was evaluated once on the holdout; the post hoc demonstration of Section 2.14 scored every pipeline on the holdout solely to quantify the optimism that selecting on the holdout would have produced.

This structure reduces a common source of optimistic bias [5] but does not turn the holdout sample into external validation. Its labels had been visible during earlier development, and all finalist evaluations in the present design used the same holdout sample. The resulting estimates are therefore conditional and internal rather than confirmatory. Confirmatory validation would require freezing the complete procedure and applying it once to data from a new hospital, time period, or prospectively collected cohort.

### 4.4 PCA and original-feature selection

PCA and original-feature selection achieved similar discrimination but differed markedly in interpretability. PCA captured variation in the expanded encoded matrix, where each one-hot level carried unit variance, and required 27 components to explain at least 95% of the variance. Each component combined multiple laboratory measurements and categorical levels and could not generally be interpreted as an individual clinical measurement [11]. Supervised selection, by contrast, retained eight named fields. The core of that panel was stable: serum creatinine, hemoglobin, packed cell volume, red blood cell count, and specific gravity were selected in every imputed training fold across four seeds, with a Nogueira stability index of 0.87–0.90 (Section 3.9). This does not establish those fields as a universally optimal clinical panel: selection dependent on the outcome was repeated within folds, and did not consider acquisition cost, fairness, or temporal availability.

The main analysis treated specific gravity, albumin, and sugar as nominal levels, in accordance with the repository metadata, although all these fields are clinically graded. Coding them as ordered integers in the prespecified sensitivity analysis lowered the repeated-cross-validation accuracy of the all-original-feature Extra Trees pipeline by 0.0071 (fold-difference range, −0.0473 to +0.0116; 10 of 15 folds unchanged) and left its ROC area at 1.0000. The difference was smaller than the pipeline's fold-to-fold standard deviation (0.0133), so the choice between nominal and ordered coding did not change the ranking or the conclusions.

The distinction matters for reproducibility. A reduced set of named variables is the result of feature selection, whereas PCA produces weighted component scores. Treating them as interchangeable obscures both the statistical operation and the clinical meaning of the resulting model.

### 4.5 Missingness and complete-case selection

Complete-case deletion generated a ceiling of exactly tied cross-validation scores while removing 60.5% of records. Because missingness was strongly associated with the outcome, deletion did more than reduce sample size. The complete-case development sample contained 81 non-CKD and 31 CKD records, while its holdout subset contained 34 non-CKD and 12 CKD records, making non-CKD the majority class. The complete-case branch, therefore, represented a different prediction problem in a selected subgroup with more complete measurements and easier class separation. Prevalence-dependent metrics such as accuracy, precision, precision-recall area, and Brier score cannot be compared directly with those from the imputed analyses.

Single imputation also has limitations. It does not propagate uncertainty in missing values, and neighbor-based distances may become unstable when many measurements are absent. Nevertheless, imputation retained the full analytical population and allowed evaluation of every holdout record. Missingness itself carried outcome information in this dataset: CKD records lacked about five times as many fields as non-CKD records, and the missing-value indicators alone discriminated the classes with a ROC area of 0.85–0.88. Any pipeline that imputes within folds—including the multiple-imputation analysis, which changed accuracy by at most 0.006—may still benefit from this pattern, which is unlikely to transfer to settings where tests are ordered differently. Future studies should confirm that all measurements are available at the intended prediction time and should establish whether the missingness pattern is part of the clinical signal or an artifact of data collection.

### 4.6 Clinical interpretation

Specific gravity, albumin, hemoglobin, packed cell volume, serum creatinine, red blood cell count, and hypertension are clinically plausible correlates of CKD status and appeared in the selected feature set, the grouped SHAP ranking, or the ablation analysis. This plausibility supports face validity but does not demonstrate prediction of incident disease or improved clinical decision-making. The UCI outcome is a diagnostic class in cross-sectional records, and some predictors may be consequences or components of an already apparent CKD presentation. The findings therefore concern classification within this dataset, not early detection before clinical recognition. Consistent with this interpretation, removing the renal-function, urinary, and hematologic fields reduced cross-validation accuracy from 0.99 to 0.89, and the demographic, metabolic, and comorbidity fields alone yielded 0.84 (Section 3.9); the near-perfect ROC areas appear to depend largely on fields that are consequences of, or definitional for, established CKD. Because the records carry no timestamps, whether each measurement preceded the diagnosis cannot be verified; Supplementary Table S6 classifies the 24 fields by their expected temporal relation to CKD recognition.

SHAP values describe how the fitted model distributes attribution among correlated, transformed predictors [15]. They depend on preprocessing, the reference distribution, feature dependence, and - under *k*-nearest-neighbor imputation - the values of other variables used to impute a field. Ablation measures performance after retraining without a field; it does not estimate the biological effect of changing that measurement. Agreement between the two approaches may aid interpretation but cannot replace causal or clinical reasoning.

### 4.7 Naive versus controlled analysis

Fitting preprocessing on all rows before cross-validation, a common form of leakage in small-sample studies [5], changed mean accuracy by less than 0.006 in this dataset but altered the ranking of pipelines ($\rho = 0.952$) and the identity of the top-ranked configuration, and selecting among the retained-sample pipelines by holdout accuracy would have reported a perfect score. In a dataset this separable, leakage therefore inflates the apparent precision of model comparison more than the level of performance: the controlled design did not lower the headline estimates, but it prevented the reporting of a spuriously identified winner. Together with the nested-resampling optimism of about 0.01 and the predictor-removal results, the sensitivity analyses suggest that the near-perfect performance is driven largely by characteristics of the data—chiefly clinically proximal laboratory and urinalysis variables, with CKD-associated missingness as an additional informative signal—and only marginally by the algorithm, the representation, or the missing-data strategy.

## 5. Strengths and Limitations

The study's strengths include schema-level inspection of the source data; deterministic repair of a shifted record; explicit normalization of CKD labels; prespecified variable types; a fixed stratified partition with persisted row assignments; and training-fold estimation of all learned preprocessing. Additional strengths include comparison of nine classifiers, repeated cross-validation on persisted folds, tie-aware selection, a single frozen holdout evaluation with conditional bootstrap intervals, calibration assessment using separated binning schemes, and complementary grouped SHAP, ablation, and ordinal-encoding analyses.

The principal limitation is a small sample. Four hundred records cannot establish clinical performance across populations, and strong class separation leaves little room for meaningful algorithm comparison; many configurations are exactly tied. The dataset came from a single historical source and lacked an external center, temporal validation cohort, prospective protocol, subgroup sizes adequate for fairness analysis, and documentation confirming that every predictor preceded the intended prediction time. The holdout is a locked historical internal sample whose labels had been visible during earlier development. Hyperparameters were fixed rather than optimized through nested resampling. Although this favored transparency, it may not have maximized each model. Class weighting was applied only to Random Forest and Extra Trees, meaning that comparisons reflected both algorithm choice and model-specific imbalance handling. A post hoc analysis that applied the same weighting policy to every classifier changed accuracy by at most 0.0044 per algorithm and, under uniform balanced weights, left the display representatives unchanged; removing weighting from every classifier moved a median/mode Extra Trees pipeline to the top of three scopes with the main representatives second, so the inconsistency affected which of two nearly tied configurations was displayed rather than the comparison itself. Evaluating 81 pipelines introduced multiplicity despite the exclusion of holdout performance from ranking. Nested resampling quantified the resulting optimism at up to about 0.01 in accuracy and showed that the selected pipeline varied across outer folds. Repeated-fold scores were dependent, so their empirical percentile ranges were not population confidence intervals. The conditional bootstrap did not propagate uncertainty from preprocessing, feature selection, training, or selection. The sensitivity analyses added after the main analysis had been frozen were post hoc, used the development folds except where stated (the missingness description, the selection-stability analysis, the all-record naive comparison, and the holdout demonstration), and cannot substitute for external or multi-center validation, which was not available. Calibration estimates were unstable because the holdout contained only 120 observations and few occupied bins. Decision-curve analysis, clinically meaningful thresholds, cost assessment, and prospective impact evaluation were not possible.

## 6. Conclusions

The leakage-controlled reproduction preserved near-perfect internal performance on the UCI CKD dataset but found only marginal, seed-dependent differences between PCA and the best original-feature approaches, although PCA improved XGBoost as in the source study; several pipelines were exactly or nearly tied. Nested resampling of the complete selection procedure, multiple imputation, uniform class weighting, and feature-selection stability analyses did not change this picture, whereas removing the strongly diagnostic laboratory and urinalysis fields reduced accuracy substantially and the missingness pattern alone was informative, indicating that the near-perfect discrimination appears to be driven largely by characteristics of the dataset, particularly clinically proximal laboratory and urinalysis variables, with informative missingness as an additional signal. Supervised selection of eight named clinical fields offered a compact and interpretable representation whose core was stable across resamples. By contrast, complete-case deletion generated deceptively strong cross-validation results after removing most records and reversing the class balance, and should be interpreted only as a sensitivity analysis. The findings favor strict separation of model development and evaluation, explicit reporting of ties, and clear distinction among discrimination, calibration, feature selection, and PCA. They do not justify clinical deployment. The complete procedure requires evaluation in larger, external, temporally appropriate, and prospectively governed cohorts.

## Acknowledgments

The authors thank the UCI Machine Learning Repository and the original dataset contributors. This study was conducted within the Molecular Basis of Disease Undergraduate Program at Georgia State University.

## Code and Data Availability

The UCI Chronic Kidney Disease dataset is publicly available from the UCI Machine Learning Repository [4]. The complete analysis code, the persisted split registries, fold-level scores, selection manifest, holdout evaluation, calibration, ablation, sensitivity, and SHAP tables, the evidence ledger linking every main-analysis value in this article to its source file, and the scripts that regenerate the main-analysis figures and tables have been prepared for release with exact dependency versions (run identifier run_2026-09-28_seed12; repository URL/DOI to be inserted). The post hoc sensitivity analyses are included in the same release as a separately hash-verified stage.

## Supplementary Material

*Supplementary Figure S1. Paired leave-one-feature-out ablation for the all-original-feature Extra Trees pipeline with numeric k-nearest-neighbor imputation. Bars show the mean paired change in repeated-cross-validation accuracy after removal of each of the 24 fields; error bars span the 2.5th–97.5th percentiles of the 15 fold-level differences. Figure 5 interprets a single development-sample fit of the same configuration.*

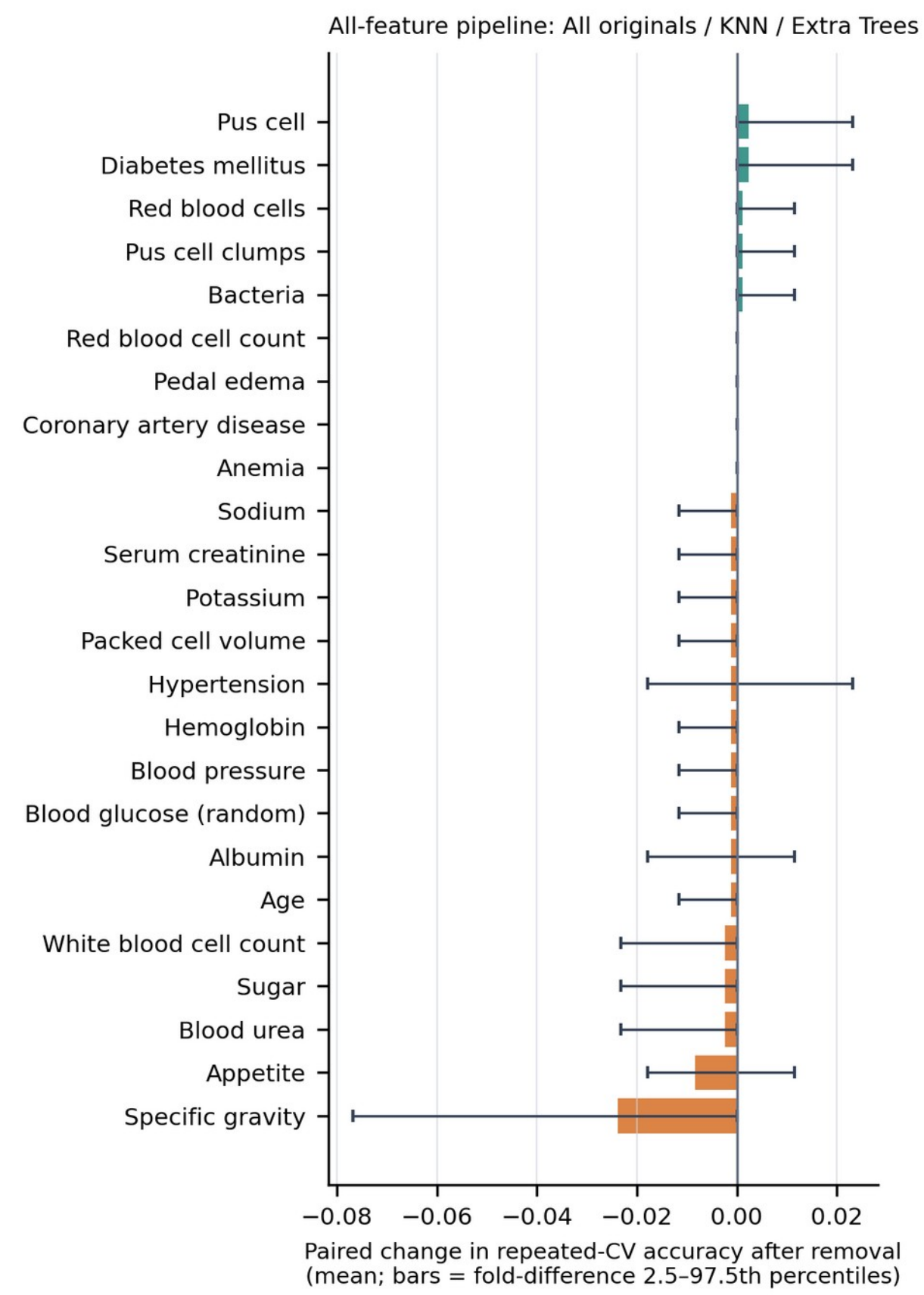

Supplementary Table S1. Seed-sensitivity replication of the prespecified selection rule

| Selection scope, seed, and outcome of the selection rule | Repeated cross-validation of the display representative | Holdout (single evaluation) of the display representative |
|---|---|---|
| All originals (retained sample); seed 12 (main analysis); display representative: All originals; KNN; Extra Trees; tied at top: 1; main-analysis representative: rank 1 (tied first) | Accuracy 0.9905 ± 0.0133; ROC-AUC 1.0000 | *n* = 120; accuracy 0.9750; ROC-AUC 1.0000; Brier 0.0149 |
| All originals (retained sample); seed 42; display representative: All originals; median/mode; SVM; tied at top: 1; main-analysis representative: rank 2 | Accuracy 0.9905 ± 0.0133; ROC-AUC 1.0000 | *n* = 120; accuracy 0.9917; ROC-AUC 0.9997; Brier 0.0073 |
| All originals (retained sample); seed 39; display representative: All originals; median/mode; SVM; tied at top: 1; main-analysis representative: rank 4 | Accuracy 0.9917 ± 0.0133; ROC-AUC 1.0000 | *n* = 120; accuracy 0.9917; ROC-AUC 0.9997; Brier 0.0061 |
| All originals (retained sample); seed 55; display representative: All originals; median/mode; Extra Trees; tied at top: 1; main-analysis representative: rank 2 | Accuracy 0.9917 ± 0.0114; ROC-AUC 1.0000 | *n* = 120; accuracy 0.9750; ROC-AUC 1.0000; Brier 0.0182 |
| PCA ≥95% (retained sample); seed 12 (main analysis); display representative: PCA ≥95%; median/mode; Random Forest; tied at top: 1; main-analysis representative: rank 1 (tied first) | Accuracy 0.9940 ± 0.0087; ROC-AUC 1.0000 | *n* = 120; accuracy 0.9917; ROC-AUC 1.0000; Brier 0.0053 |

| Selection scope, seed, and outcome of the selection rule | Repeated cross-validation of the display representative | Holdout (single evaluation) of the display representative |
|---|---|---|
| PCA ≥95% (retained sample); seed 42; display representative: PCA ≥95%; median/mode; Extra Trees; tied at top: 1; main-analysis representative: rank 6 | Accuracy 0.9893 ± 0.0132; ROC-AUC 1.0000 | *n* = 120; accuracy 0.9917; ROC-AUC 1.0000; Brier 0.0090 |
| PCA ≥95% (retained sample); seed 39; display representative: PCA ≥95%; median/mode; Random Forest; tied at top: 1; main-analysis representative: rank 1 (tied first) | Accuracy 0.9940 ± 0.0087; ROC-AUC 1.0000 | *n* = 120; accuracy 0.9917; ROC-AUC 1.0000; Brier 0.0053 |
| PCA ≥95% (retained sample); seed 55; display representative: PCA ≥95%; median/mode; Extra Trees; tied at top: 1; main-analysis representative: rank 6 | Accuracy 0.9964 ± 0.0074; ROC-AUC 1.0000 | *n* = 120; accuracy 0.9750; ROC-AUC 0.9985; Brier 0.0189 |
| Selected originals (retained sample); seed 12 (main analysis); display representative: Selected originals; KNN; Extra Trees; tied at top: 1; main-analysis representative: rank 1 (tied first) | Accuracy 0.9869 ± 0.0218; ROC-AUC 1.0000 | *n* = 120; accuracy 0.9833; ROC-AUC 1.0000; Brier 0.0142 |
| Selected originals (retained sample); seed 42; display representative: Selected originals; KNN; Extra Trees; tied at top: 1; main-analysis representative: rank 1 (tied first) | Accuracy 0.9952 ± 0.0082; ROC-AUC 1.0000 | *n* = 120; accuracy 0.9750; ROC-AUC 0.9994; Brier 0.0160 |
| Selected originals (retained sample); seed 39; display representative: Selected originals; KNN; Extra Trees; tied at top: 1; main-analysis representative: rank 1 (tied first) | Accuracy 0.9881 ± 0.0129; ROC-AUC 1.0000 | *n* = 120; accuracy 0.9833; ROC-AUC 0.9997; Brier 0.0087 |
| Selected originals (retained sample); seed 55; display representative: Selected originals; median/mode; Extra Trees; tied at top: 1; main-analysis representative: rank 2 | Accuracy 0.9810 ± 0.0126; ROC-AUC 1.0000 | *n* = 120; accuracy 0.9750; ROC-AUC 1.0000; Brier 0.0134 |
| Complete-case sensitivity; seed 12 (main analysis); display representative: All originals; complete case; Extra Trees; tied at top: 8; main-analysis representative: rank 1 (tied first) | Accuracy 1.0000 ± 0.0000; ROC-AUC 1.0000 | *n* = 46; accuracy 1.0000; ROC-AUC 1.0000; Brier 0.0030 |
| Complete-case sensitivity; seed 42; display representative: All originals; complete case; Extra Trees; tied at top: 15; main-analysis representative: rank 1 (tied first) | Accuracy 1.0000 ± 0.0000; ROC-AUC 1.0000 | *n* = 50; accuracy 0.9800; ROC-AUC 1.0000; Brier 0.0076 |
| Complete-case sensitivity; seed 39; display representative: All originals; complete case; Extra Trees; tied at top: 13; main-analysis representative: rank 1 (tied first) | Accuracy 1.0000 ± 0.0000; ROC-AUC 1.0000 | *n* = 48; accuracy 1.0000; ROC-AUC 1.0000; Brier 0.0052 |
| Complete-case sensitivity; seed 55; display representative: All originals; complete case; Extra Trees; tied at top: 8; main-analysis representative: rank 1 (tied first) | Accuracy 1.0000 ± 0.0000; ROC-AUC 1.0000 | *n* = 46; accuracy 1.0000; ROC-AUC 1.0000; Brier 0.0006 |
| Non-PCA tree models (retained sample); seed 12 (main analysis); display representative: All originals; KNN; Extra Trees; tied at top: 1; main-analysis representative: rank 1 (tied first) | Accuracy 0.9905 ± 0.0133; ROC-AUC 1.0000 | *n* = 120; accuracy 0.9750; ROC-AUC 1.0000; Brier 0.0149 |
| Non-PCA tree models (retained sample); seed 42; display representative: Selected originals; KNN; Extra Trees; tied at top: 1; main-analysis representative: rank 2 | Accuracy 0.9952 ± 0.0082; ROC-AUC 1.0000 | *n* = 120; accuracy 0.9750; ROC-AUC 0.9994; Brier 0.0160 |
| Non-PCA tree models (retained sample); seed 39; display representative: Selected originals; KNN; Extra Trees; tied at top: 1; main-analysis representative: rank 4 | Accuracy 0.9881 ± 0.0129; ROC-AUC 1.0000 | *n* = 120; accuracy 0.9833; ROC-AUC 0.9997; Brier 0.0087 |
| Non-PCA tree models (retained sample); seed 55; display representative: All originals; median/mode; Extra Trees; tied at top: 1; main-analysis representative: rank 2 | Accuracy 0.9917 ± 0.0114; ROC-AUC 1.0000 | *n* = 120; accuracy 0.9750; ROC-AUC 1.0000; Brier 0.0182 |

*RCV, repeated cross-validation (five folds, three repeats) on the development sample of the indicated seed; the main analysis used seed 12. Each seed redraws the stratified 70:30 partition, so holdout samples differ between rows and are not independent. "Main-analysis representative: rank" is the dense rank, within the same scope and seed, of the configuration displayed in the main analysis. Holdout metrics of the replications are single observed values without bootstrap intervals.*

Supplementary Table S2. Missingness by outcome class for the 24 predictors (*N* = 400; CKD *n* = 250, non-CKD *n* = 150)

| Field (abbreviation) | Type | Missing, all records, *n* (%) | Missing, CKD, *n* (%) | Missing, non-CKD, *n* (%) | Ratio of missingness rates, CKD/non-CKD |
|---|---|---|---|---|---|
| Age (age) | Numeric | 9 (2.3%) | 8 (3.2%) | 1 (0.7%) | 4.8 |
| Blood pressure (bp) | Numeric | 12 (3.0%) | 10 (4.0%) | 2 (1.3%) | 3.0 |
| Specific gravity (sg) | Categorical | 47 (11.8%) | 42 (16.8%) | 5 (3.3%) | 5.0 |
| Albumin (al) | Categorical | 46 (11.5%) | 41 (16.4%) | 5 (3.3%) | 4.9 |
| Sugar (su) | Categorical | 49 (12.3%) | 44 (17.6%) | 5 (3.3%) | 5.3 |
| Red blood cells (morphology) (rbc) | Categorical | 152 (38.0%) | 143 (57.2%) | 9 (6.0%) | 9.5 |
| Pus cells (pc) | Categorical | 65 (16.3%) | 56 (22.4%) | 9 (6.0%) | 3.7 |

| Field (abbreviation) | Type | Missing, all records, *n* (%) | Missing, CKD, *n* (%) | Missing, non-CKD, *n* (%) | Ratio of missingness rates, CKD/non-CKD |
|---|---|---|---|---|---|
| Pus cell clumps (pcc) | Categorical | 4 (1.0%) | 0 (0.0%) | 4 (2.7%) | — |
| Bacteria (ba) | Categorical | 4 (1.0%) | 0 (0.0%) | 4 (2.7%) | — |
| Blood glucose (random) (bgr) | Numeric | 44 (11.0%) | 38 (15.2%) | 6 (4.0%) | 3.8 |
| Blood urea (bu) | Numeric | 19 (4.8%) | 13 (5.2%) | 6 (4.0%) | 1.3 |
| Serum creatinine (sc) | Numeric | 17 (4.3%) | 12 (4.8%) | 5 (3.3%) | 1.4 |
| Sodium (sod) | Numeric | 87 (21.8%) | 82 (32.8%) | 5 (3.3%) | 9.8 |
| Potassium (pot) | Numeric | 88 (22.0%) | 83 (33.2%) | 5 (3.3%) | 10.0 |
| Hemoglobin (hemo) | Numeric | 52 (13.0%) | 46 (18.4%) | 6 (4.0%) | 4.6 |
| Packed cell volume (pcv) | Numeric | 71 (17.8%) | 67 (26.8%) | 4 (2.7%) | 10.1 |
| White blood cell count (wbcc) | Numeric | 106 (26.5%) | 99 (39.6%) | 7 (4.7%) | 8.5 |
| Red blood cell count (rbcc) | Numeric | 131 (32.8%) | 124 (49.6%) | 7 (4.7%) | 10.6 |
| Hypertension (htn) | Categorical | 2 (0.5%) | 0 (0.0%) | 2 (1.3%) | — |
| Diabetes mellitus (dm) | Categorical | 2 (0.5%) | 0 (0.0%) | 2 (1.3%) | — |
| Coronary artery disease (cad) | Categorical | 2 (0.5%) | 0 (0.0%) | 2 (1.3%) | — |
| Appetite (appet) | Categorical | 1 (0.3%) | 0 (0.0%) | 1 (0.7%) | — |
| Pedal edema (pe) | Categorical | 1 (0.3%) | 0 (0.0%) | 1 (0.7%) | — |
| Anemia (ane) | Categorical | 1 (0.3%) | 0 (0.0%) | 1 (0.7%) | — |

*CKD records lacked a mean of 3.63 fields (median, 3; 36.0% with five or more missing) and non-CKD records 0.69 (median, 0; 4.0%); 17.2% of CKD and 76.7% of non-CKD records were complete. Classifiers fitted to the 24 binary missing-value indicators alone on the persisted development folds: logistic regression, accuracy 0.8131 ± 0.0301 and ROC area 0.8540 ± 0.0363; Extra Trees, accuracy 0.8202 ± 0.0305 and ROC area 0.8779 ± 0.0293. — indicates that no CKD record lacked the field (ratio 0).*

Supplementary Table S3. Multiple imputation compared with single imputation for the retained-sample display representatives (20 imputations, identical folds)

| Pipeline | Single-imputation accuracy | Pooled MI accuracy, mean ± SD | Δ accuracy, MI − single (paired fold range) | Pooled MI ROC area | Δ ROC area | Between-imputation SD of fold accuracy |
|---|---|---|---|---|---|---|
| All originals; KNN; Extra Trees | 0.9905 | 0.9905 ± 0.0133 | 0.0000 (0.0000 to 0.0000) | 1.0000 | 0.0000 | 0.0014 |
| PCA ≥95%; median/mode; Random Forest | 0.9940 | 0.9929 ± 0.0091 | −0.0012 (−0.0116 to 0.0000) | 0.9999 | −0.0001 | 0.0109 |
| Selected originals; KNN; Extra Trees | 0.9869 | 0.9810 ± 0.0208 | −0.0060 (−0.0295 to +0.0116) | 0.9998 | −0.0002 | 0.0100 |

*MI, multiple imputation; SD, standard deviation. Numeric predictors were imputed 20 times within each training fold by chained equations (Bayesian ridge regression with posterior sampling); categorical predictors kept the training-fold mode, so the PCA pipeline's median/mode reference differs from the MI variant only in its numeric imputer. Validation-fold probabilities were averaged across imputations before scoring. Fold ranges are the 2.5th–97.5th percentiles of the 15 paired fold differences. Differences were computed before rounding.*

Supplementary Table S4. Stability of the eight-field supervised selection across 60 training folds (four seeds × 15 folds)

| Missing-data strategy | Nogueira stability index (bootstrap 2.5th–97.5th percentiles) | Nogueira index, seed 12 only | Mean pairwise Jaccard similarity | Distinct panels | Fields selected in all 60 folds | Fields selected in a majority of folds |
|---|---|---|---|---|---|---|
| Median/mode | 0.902 (0.873–0.936) | 0.952 | 0.885 | 6 | sc, hemo, pcv, rbcc, sg | sc, sod, hemo, pcv, rbcc, sg, al, htn |
| *k*-nearest neighbor | 0.875 (0.858–0.896) | 0.911 | 0.853 | 6 | sc, hemo, pcv, rbcc, sg, al | sc, sod, hemo, pcv, rbcc, sg, al, htn |
| Complete case | 0.869 (0.848–0.897) | 0.886 | 0.847 | 7 | sc, hemo, pcv, rbcc, al | bu, sc, hemo, pcv, rbcc, sg, al, htn |

*The selector retained the eight fields with the highest training-fold mutual information with the outcome (Section 2.6). Field abbreviations follow the UCI repository: sc, serum creatinine; bu, blood urea; hemo, hemoglobin; pcv, packed cell volume; rbcc, red blood cell count; sg, specific gravity; al, albumin; htn, hypertension; sod, sodium. The bootstrap interval resamples the 60 selection sets 2,000 times. Selection frequencies of every field are shown in Supplementary Figure S3.*

Supplementary Table S5. Repeated-cross-validation performance after removal of diagnostic predictors (development folds only)

| Removal or retention set | Fields retained | All originals; KNN; Extra Trees: accuracy (Δ vs. reference; paired fold range) | ROC area (Δ) | Selected originals; KNN; Extra Trees: accuracy (Δ vs. reference; paired fold range) | ROC area (Δ) |
|---|---|---|---|---|---|
| All 24 fields (reference) | 24 | 0.9905 (0.0000; 0.0000 to 0.0000) | 1.0000 (0.0000) | 0.9869 (0.0000; 0.0000 to 0.0000) | 1.0000 (0.0000) |
| Remove renal-function markers | 22 | 0.9893 (−0.0012; −0.0232 to +0.0116) | 1.0000 (0.0000) | 0.9845 (−0.0024; −0.0357 to +0.0179) | 0.9990 (−0.0010) |
| Remove urinary damage markers | 18 | 0.9560 (−0.0345; −0.0830 to 0.0000) | 0.9956 (−0.0044) | 0.9440 (−0.0429; −0.0714 to −0.0063) | 0.9880 (−0.0120) |
| Remove hematologic fields | 19 | 0.9929 (+0.0024; 0.0000 to +0.0232) | 0.9983 (−0.0017) | 0.9821 (−0.0048; −0.0295 to +0.0179) | 0.9954 (−0.0046) |
| Remove electrolytes | 22 | 0.9905 (0.0000; 0.0000 to 0.0000) | 1.0000 (0.0000) | 0.9869 (0.0000; −0.0179 to +0.0179) | 1.0000 (0.0000) |
| Remove metabolic and comorbidity fields | 19 | 0.9869 (−0.0036; −0.0295 to +0.0295) | 0.9999 (−0.0001) | 0.9821 (−0.0048; −0.0357 to +0.0411) | 0.9988 (−0.0012) |
| Remove symptoms and signs | 22 | 0.9798 (−0.0107; −0.0295 to 0.0000) | 0.9978 (−0.0022) | 0.9869 (0.0000; 0.0000 to 0.0000) | 1.0000 (0.0000) |
| Remove renal-function and urinary markers | 16 | 0.9583 (−0.0321; −0.0830 to +0.0054) | 0.9921 (−0.0079) | 0.9357 (−0.0512; −0.1071 to +0.0054) | 0.9790 (−0.0210) |
| Remove renal-function, urinary, and hematologic fields | 11 | 0.8917 (−0.0988; −0.1545 to −0.0598) | 0.9794 (−0.0206) | 0.8917 (−0.0952; −0.1607 to −0.0482) | 0.9818 (−0.0182) |

| Removal or retention set | Fields retained | All originals; KNN; Extra Trees: accuracy (Δ vs. reference; paired fold range) | ROC area (Δ) | Selected originals; KNN; Extra Trees: accuracy (Δ vs. reference; paired fold range) | ROC area (Δ) |
|---|---|---|---|---|---|
| Keep only serum creatinine and blood urea | 2 | 0.8357 (−0.1548; −0.2259 to −0.0893) | 0.9357 (−0.0643) | 0.8357 (−0.1512; −0.2375 to −0.0893) | 0.9357 (−0.0643) |
| Keep only serum creatinine and albumin | 2 | 0.8750 (−0.1155; −0.2795 to −0.0357) | 0.9519 (−0.0481) | 0.8750 (−0.1119; −0.2616 to −0.0241) | 0.9519 (−0.0481) |
| Keep only demographic, vital-sign, metabolic, and comorbidity fields | 7 | 0.8369 (−0.1536; −0.2196 to −0.0955) | 0.9594 (−0.0406) | 0.8369 (−0.1500; −0.2196 to −0.1071) | 0.9594 (−0.0406) |
| Remove top 1 grouped-SHAP field (htn) | 23 | 0.9893 (−0.0012; −0.0179 to +0.0232) | 0.9996 (−0.0004) | 0.9869 (0.0000; −0.0179 to +0.0232) | 1.0000 (0.0000) |
| Remove top 2 grouped-SHAP fields (+ al) | 22 | 0.9821 (−0.0083; −0.0357 to +0.0116) | 0.9995 (−0.0005) | 0.9750 (−0.0119; −0.0411 to +0.0116) | 0.9988 (−0.0012) |
| Remove top 3 grouped-SHAP fields (+ dm) | 21 | 0.9798 (−0.0107; −0.0357 to +0.0179) | 0.9977 (−0.0023) | 0.9595 (−0.0274; −0.0830 to 0.0000) | 0.9944 (−0.0056) |
| Remove top 4 grouped-SHAP fields (+ sg) | 20 | 0.9631 (−0.0274; −0.0830 to +0.0116) | 0.9912 (−0.0088) | 0.9393 (−0.0476; −0.1009 to 0.0000) | 0.9884 (−0.0116) |
| Remove top 5 grouped-SHAP fields (+ appet) | 19 | 0.9548 (−0.0357; −0.0946 to +0.0116) | 0.9885 (−0.0115) | 0.9417 (−0.0452; −0.1009 to 0.0000) | 0.9900 (−0.0100) |
| Remove top 6 grouped-SHAP fields (+ hemo) | 18 | 0.9333 (−0.0571; −0.1188 to −0.0179) | 0.9897 (−0.0103) | 0.9333 (−0.0536; −0.0893 to −0.0179) | 0.9904 (−0.0096) |
| Remove top 7 grouped-SHAP fields (+ pcv) | 17 | 0.9179 (−0.0726; −0.1250 to −0.0357) | 0.9892 (−0.0108) | 0.9143 (−0.0726; −0.1188 to −0.0241) | 0.9830 (−0.0170) |
| Remove top 8 grouped-SHAP fields (+ pe) | 16 | 0.9071 (−0.0833; −0.1482 to −0.0357) | 0.9861 (−0.0139) | 0.9131 (−0.0738; −0.1250 to −0.0357) | 0.9784 (−0.0216) |

*Δ, mean paired difference from the same pipeline with all 24 fields over the 15 persisted folds; fold ranges are the 2.5th–97.5th percentiles of the paired differences. For the selected-feature pipeline, supervised selection of up to eight fields was repeated within each training fold on the retained fields. Grouped-SHAP order (Section 3.7): htn, al, dm, sg, appet, hemo, pcv, pe. Group membership is given in Supplementary Table S6. Differences were computed before rounding.*

Supplementary Table S6. Clinical grouping of the 24 predictors and their expected temporal relation to CKD diagnosis

| Group | Fields | Expected relation to the CKD diagnosis |
|---|---|---|
| Renal-function markers | Serum creatinine (sc); Blood urea (bu) | Definitional: serum creatinine underlies eGFR-based staging; urea rises as filtration falls. Measured at or after presentation. |
| Urinary damage markers | Albumin (al); Specific gravity (sg); Red blood cells (morphology) (rbc); Pus cells (pc); Pus cell clumps (pcc); Bacteria (ba) | Definitional or diagnostic work-up: albuminuria and urinalysis findings are KDIGO markers of kidney damage. |
| Hematologic fields | Hemoglobin (hemo); Packed cell volume (pcv); Red blood cell count (rbcc); White blood cell count (wbcc); Anemia (ane) | Consequences of established disease (renal anemia) or general hematology ordered during the work-up. |
| Electrolytes | Sodium (sod); Potassium (pot) | Consequences of reduced excretion; measured with the renal panel. |
| Metabolic and comorbidity fields | Blood glucose (random) (bgr); Sugar (su); Hypertension (htn); Diabetes mellitus (dm); Coronary artery disease (cad) | Risk factors and comorbidities that can precede the diagnosis. |
| Symptoms and signs | Appetite (appet); Pedal edema (pe) | Late uremic symptoms and signs; typically follow established disease. |
| Demographic and vital-sign fields | Age (age); Blood pressure (bp) | Available before diagnosis; blood pressure is both a cause and a consequence. |

*Groupings follow the KDIGO definition of CKD by markers of kidney damage and reduced glomerular filtration [2]. The UCI records carry no timestamps or encounter structure, so the temporal relation is the expected clinical sequence rather than an empirical finding; it motivates the removal sets of Supplementary Table S5.*

Supplementary Table S7. Effect of a uniform class-weighting policy on each algorithm across the six retained-sample designs

| Algorithm | Weighting applied under balanced policy | Accuracy, no weighting | Accuracy, balanced weights | Δ accuracy, balanced − none, mean (min to max over designs) | Δ ROC area, mean |
|---|---|---|---|---|---|
| Logistic Regression | Yes | 0.9677 | 0.9637 | −0.0040 (−0.0071 to 0.0000) | +0.0001 |
| Decision Tree | Yes | 0.9687 | 0.9730 | +0.0044 (0.0000 to +0.0095) | +0.0031 |
| Random Forest | Yes | 0.9833 | 0.9853 | +0.0020 (−0.0012 to +0.0048) | +0.0002 |
| SVM | Yes | 0.9829 | 0.9823 | −0.0006 (−0.0012 to 0.0000) | −0.0003 |
| KNN | No (not supported) | 0.9548 | 0.9548 | 0.0000 (0.0000 to 0.0000) | 0.0000 |
| AdaBoost | Yes | 0.9665 | 0.9621 | −0.0044 (−0.0095 to 0.0000) | +0.0003 |
| Gradient Boosting | Yes | 0.9813 | 0.9827 | +0.0014 (−0.0012 to +0.0036) | +0.0002 |
| Extra Trees | Yes | 0.9907 | 0.9885 | −0.0022 (−0.0071 to 0.0000) | 0.0000 |
| XGBoost | Yes | 0.9778 | 0.9776 | −0.0002 (−0.0024 to +0.0024) | +0.0001 |

*Each algorithm was evaluated in six retained-sample designs (three representations × two imputation strategies) on the persisted folds under two uniform policies: no class weighting for any classifier, and balanced class weights passed as training-fold sample weights to every classifier that accepts them. Under the balanced policy, reapplying the selection rule reproduced all six display representatives of the main analysis; under no weighting, the all-original, selected-feature, and non-PCA tree scopes selected the median/mode Extra Trees pipeline and the main representatives ranked second (Spearman ρ between retained-sample accuracy rankings and the main analysis, 0.986 and 0.987). Pipelines whose policy matched the main analysis reproduced its fold scores exactly. Differences were computed before rounding.*

Supplementary Table S8. Naive global preprocessing compared with the leakage-controlled design on identical folds

| Analysis | *n* | Top-ranked retained-sample pipeline | Its accuracy | Mean accuracy of 54 retained-sample pipelines | Pipelines with ROC area 1.0000 | Complete-case pipelines tied at the top | Spearman ρ vs. controlled development ranking |
|---|---|---|---|---|---|---|---|
| Controlled (fold-wise preprocessing), development sample | 280 | PCA; median/mode; Random Forest | 0.9940 | 0.9748 | 8 | 8 | 1.000 |

| Analysis | *n* | Top-ranked retained-sample pipeline | Its accuracy | Mean accuracy of 54 retained-sample pipelines | Pipelines with ROC area 1.0000 | Complete-case pipelines tied at the top | Spearman ρ vs. controlled development ranking |
|---|---|---|---|---|---|---|---|
| Naive (preprocessing fitted once on the development rows), development sample | 280 | PCA; KNN; Random Forest | 0.9952 | 0.9752 | 11 | 9 | 0.952 |
| Controlled, all 400 records | 400 | Selected originals; KNN; Extra Trees | 0.9925 | 0.9805 | 6 | 8 | 0.886 |
| Naive, all 400 records | 400 | PCA; KNN; Extra Trees | 0.9908 | 0.9792 | 6 | 8 | 0.899 |

*In the naive analyses, imputation, scaling, supervised feature selection, and PCA were fitted once on all rows of the indicated population before cross-validation and then held fixed; the controlled analyses refitted them within every training fold. Both used the same five-fold, three-repeat folds (seed 12). Holdout demonstration: evaluating each of the 54 retained-sample controlled pipelines once on the locked holdout and reporting the best accuracy would have given 1.0000 (8 pipelines tied; 23 of the 27 complete-case pipelines also reached 1.0000 on their 46-record subset), compared with 0.9750–1.0000 for the prespecified frozen finalists.*

*Supplementary Figure S2. Nested resampling of the selection procedure compared with apparent estimates. For each scope, circles show the mean outer-fold accuracy (left) and ROC area (right) of the pipeline selected within each outer training fold, with bars spanning the 2.5th–97.5th percentiles of the 15 outer folds; squares show the selected pipeline's inner cross-validation estimate, and diamonds the main-analysis display representative evaluated on the same folds without nesting. Numbers in parentheses are candidate counts; letters identify the feature representation (A, all original fields; B, PCA; C, selected original fields), each with 18 candidates, and the non-PCA tree scope has 20 (Table 5). The complete-case branch is a different population (n = 112).*

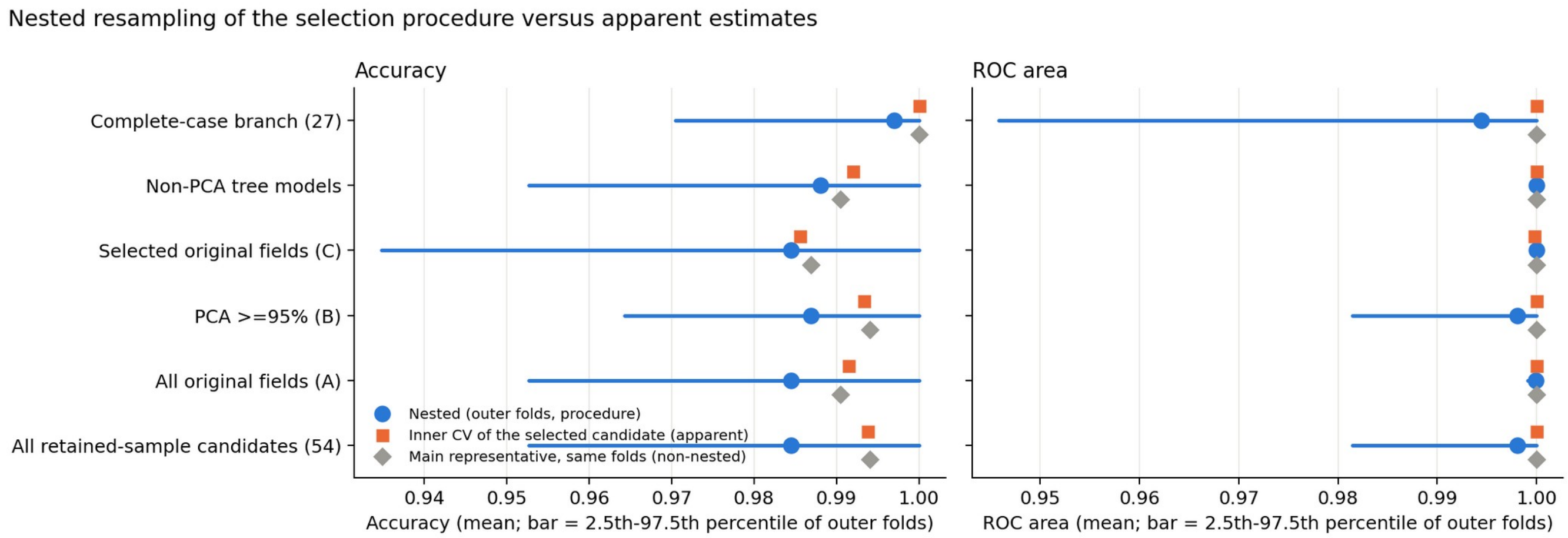

*Supplementary Figure S3. Selection frequency of each field in the eight-field supervised selector across 60 training folds (four seeds × 15 folds) under each missing-data strategy, with the Nogueira stability index and its bootstrap 2.5th–97.5th percentile interval. Field abbreviations follow the UCI repository (Supplementary Table S2).*

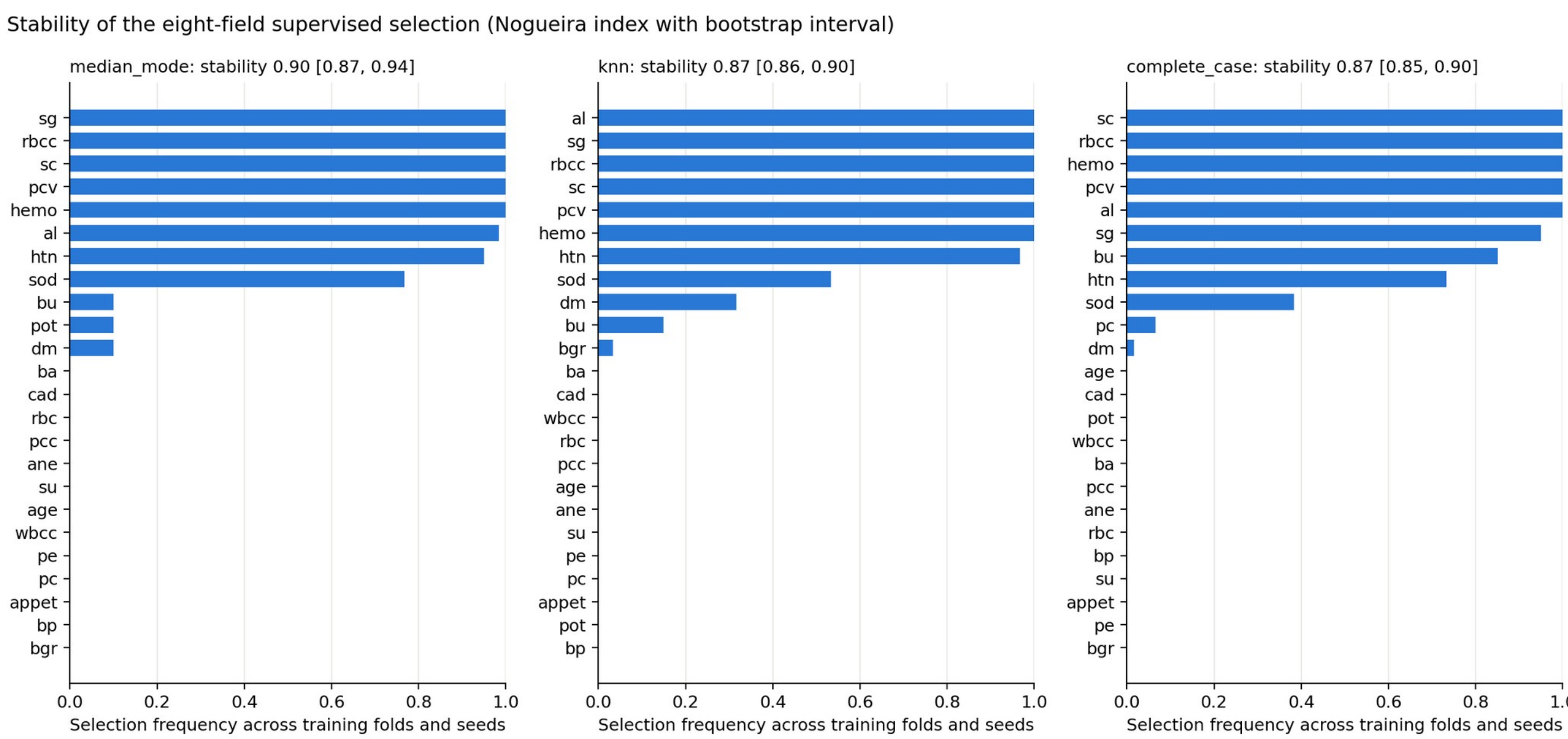